\newif\ifdoubleblind
\newif\ifaccess
\ifaccess
	\documentclass{ieeeaccess}
\else
	\documentclass[journal]{IEEEtran}
\fi

\usepackage{cite}
\usepackage{amsmath,amssymb,amsfonts}
\usepackage{algorithmic}
\usepackage{textcomp}
\DeclareMathOperator*{\argmax}{arg\,max}

\usepackage{graphicx}
\usepackage{caption,setspace}
\usepackage{tikz}
\usetikzlibrary{positioning}
\usepackage{pdfcomment}
\usepackage[caption=false,font=footnotesize]{subfig}
\usepackage[nolist]{acronym}
\usepackage{xspace}
\usepackage{url}
\usepackage{hyperref}

\usepackage{tabularx,booktabs}
\usepackage{multirow}
\usepackage{rotating}
\usepackage{wasysym}
\usepackage{fontawesome}

\usepackage{listings}
\definecolor{codegreen}{rgb}{0,0.6,0}
\definecolor{codegray}{rgb}{0.5,0.5,0.5}
\definecolor{codepurple}{rgb}{0.58,0,0.82}
\definecolor{backcolour}{rgb}{0.9020,0.9294,0.9608}
\lstdefinestyle{mystyle}{
	backgroundcolor=\color{backcolour},   
	commentstyle=\color{codegreen},
	keywordstyle=\color{magenta},
	numberstyle=\tiny\color{codegray},
	stringstyle=\color{codepurple},
	basicstyle=\ttfamily\footnotesize,
	breakatwhitespace=false,         
	breaklines=true,                 
	captionpos=b,                    
	keepspaces=true,                 
	numbers=left,                    
	numbersep=5pt,                  
	showspaces=false,                
	showstringspaces=false,
	showtabs=false,                  
	tabsize=2
}
\begin{document}

\newcommand{\paperTitle}{Client-Based Intelligence for Resource Efficient Vehicular Big Data Transfer  in Future 6G Networks}

\newcommand{\paperAuthors}{Benjamin Sliwa\thanks{The authors are with the Communication Networks Institute,TU Dortmund University, 44227 Dortmund, Germany (e-mail: \paperEmails)}
	, \emph{Student Member, IEEE}, Rick Adam, and Christian Wietfeld, \emph{Senior Member, IEEE}}
\newcommand{\paperEmails}{$\{$Benjamin.Sliwa, Rick Adam, Christian.Wietfeld$\}$@tu-dortmund.de}

\newcommand{\figurePadding}{0pt}
\newcommand{\figureTopPadding}{\figurePadding}
\newcommand{\figureBottomPadding}{\figurePadding}
\newcommand\red[1]{\colorbox{red}{\textbf{TODO: #1}}}
\newcommand\mno[1]{\emph{\ac{MNO}~#1\xspace}}

\newcommand{\boxWidth}{0.47\textwidth}
\newcommand\q{Q(\mathbf{c_{t}},a)}
\newcommand\qq{Q(\mathbf{c_{t+1}},a)}
\newcommand{\proposal}{\ac{BS-CB}\xspace}

\newcommand{\dummy}[3]
{
	\begin{figure}[b!]  
		\begin{tikzpicture}
		\node[draw,minimum height=6cm,minimum width=\columnwidth]{\LARGE #1};
		\end{tikzpicture}
		\caption{#2}
		\label{#3}
	\end{figure}
}

\newcommand{\wDummy}[3]
{
	\begin{figure*}[b!]  
		\begin{tikzpicture}
		\node[draw,minimum height=6cm,minimum width=\textwidth]{\LARGE #1};
		\end{tikzpicture}
		\caption{#2}
		\label{#3}
	\end{figure*}
}

\newcommand{\basicFig}[7]
{
	\begin{figure}[#1]  	
		\vspace{#6}
		\centering		  
		\includegraphics[width=#7\columnwidth]{#2}
		\caption{#3}
		\label{#4}
		\vspace{#5}	
	\end{figure}
}
\newcommand{\fig}[4]{\basicFig{#1}{#2}{#3}{#4}{0cm}{0cm}{1}}

\newcommand{\subfig}[3]
{
	\subfloat[#3]
	{
		\includegraphics[width=#2\textwidth]{#1}
	}
	\hfill
}

\newcommand\circled[1] % caution with using in captions: \protect \circled
{
	\tikz[baseline=(char.base)]
	{
		\node[shape=circle,draw,inner sep=1pt] (char) {#1};
	}\xspace
}
\begin{acronym}
	\acro{LTE}{Long Term Evolution}
	\acro{UE}{User Equipment}
	\acro{eNB}{evolved Node B}
	\acro{MNO}{Mobile Network Operator}
	\acro{AoI}{Age of Information}
	\acro{QoS}{Quality of Service}
	\acro{RF}{Random Forest}
	\acro{ANN}{Artificial Neural Network}
	\acro{GPR}{Gaussian Process Regression}
	\acro{SVM}{Support Vector Machine}
	\acro{M5}{M5 Regression Tree}
	\acro{NWDAF}{Network Data Analytics Function}
	\acro{WEKA}{Waikato Environment for Knowledge Analysis}
	\acro{RMSE}{Root Mean Squared Error}
	\acro{MAE}{Mean Absolute Error}
	\acro{DDNS}{Data-driven Network Simulation}
	\acro{LIMITS}{Lightweight Machine Learning for IoT Systems}
	\acro{LSTM}{Long Short-term Memory}
	\acro{TCP}{Transmission Control Protocol}
	\acro{CAT}{Channel-aware Transmission}
	\acro{pCAT}{predictive CAT}
	\acro{ML-CAT}{Machine Learning CAT}
	\acro{ML-pCAT}{Machine Learning pCAT}
	\acro{RL-CAT}{Reinforcement Learning CAT}
	\acro{RL-pCAT}{Reinforcement Learnign pCAT}
	\acro{CASTLE}{Client-side Adaptive Scheduler That minimizes Load and Energy}
	\acro{CART}{Classification and Regression Tree}
	\acro{MCS}{Modulation and Coding Scheme}
	\acro{HARQ}{Hybrid Automatic Repeat Request}
	\acro{MAC}{Medium Access Control}
	\acro{TTI}{Transmission Time Interval}
	\acro{TBS}{Transport Block Size}
	\acro{PRB}{Physical Resource Block}
	\acro{CBR}{Constant Bitrate}
	\acro{5GAA}{5G Automotive Association}
	\acro{ASU}{Arbitrary Strength Unit}
	\acro{RSSI}{Reference Signal Strength Indicator}
	\acro{RSRP}{Reference Signal Received Power}
	\acro{RSRQ}{Reference Signal Received Quality}
	\acro{SINR}{Signal-to-interference-plus-noise Ratio}
	\acro{CQI}{Channel Quality Indicator}
	\acro{TA}{Timing Advance}	
	\acro{ECDF}{Empirical Cumulative Distribution Function}
	\acro{mMTC}{massive Machine-type Communication}
	\acro{MTC}{Machine-type Communication}
	\acro{ITU}{International Telecommunication Union}
	\acro{HD}{High Definition}
	\acro{SMO}{Sequential Minimal Optimization}
	\acro{RBF}{Radial Basis Function}
	\acro{KPI}{Key Performance Indicator}
	\acro{UAV}{Unmanned Aerial Vehicle}
	\acro{ITS}{Intelligent Transportation System}
	\acro{CoAST}{Collaborative Application-Aware Scheduling of Last Mile Cellular Traffic}
	\acro{RAT}{Radio Access Technology}
	\acro{LinUCB}{Linear Upper Confidence Bound}
	\acro{BS-CB}{Black Spot-aware Contextual Bandit}
	\acro{AECC}{Automotive Edge Computing Consortium}
\end{acronym}

\acresetall
\ifaccess
\history{Date of publication xxxx 00, 0000, date of current version xxxx 00, 0000.}
\doi{10.1109/ACCESS.2017.DOI}
\title{Optimizing Cognitive Vehicular Communication Systems with Data-driven Network Simulation}
\author{\uppercase{Benjamin Sliwa}\authorrefmark{1}, \IEEEmembership{Student Member, IEEE},
	\uppercase{Christian Wietfeld}\authorrefmark{1}, \IEEEmembership{Senior Member, IEEE}}
\address[1]{Communication Networks Institute, TU Dortmund University, 44227 Dortmund, Germany}
%
% Acknowledgment
%
\tfootnote{Part of the work on this paper has been supported by Deutsche Forschungsgemeinschaft (DFG) within the Collaborative Research Center SFB 876 ``Providing Information by Resource-Constrained Analysis'', project B4.}
\markboth
{Author \headeretal: Preparation of Papers for IEEE TRANSACTIONS and JOURNALS}
{Author \headeretal: Preparation of Papers for IEEE TRANSACTIONS and JOURNALS}

\corresp{Corresponding author: Benjamin Sliwa (e-mail: benjamin.sliwa@tu-dortmund.de).}

\titlepgskip=-15pt
\else
\title{\paperTitle}

\author{\IEEEauthorblockN{{\paperAuthors}}\\
	%\IEEEauthorblockA{Communication Networks Institute,	TU Dortmund University, 44227 Dortmund, Germany\\
	%	e-mail: \paperEmails}
	}

\maketitle

\fi

\begin{tikzpicture}[remember picture, overlay]
\node[below=5mm of current page.north, text width=20cm,font=\sffamily\footnotesize,align=center] {Accepted for publication in: IEEE Transactions on Vehicular Technology\vspace{0.3cm}\\\pdfcomment[color=yellow,icon=Note]{
@Article\{Sliwa/etal/2021b,\\
  author  = \{Benjamin Sliwa and Rick Adam and Christian Wietfeld\},\\
  journal = \{IEEE Transactions on Vehicular Technology\},\\
  title   = \{Client-based intelligence for resource efficient vehicular big data transfer in future \{6G\} networks\},\\
  year    = \{2021\},\\
  month   = \{Feb\},\\
\}
}};
\node[above=5mm of current page.south, text width=15cm,font=\sffamily\footnotesize] {2021~IEEE. Personal use of this material is permitted. However, permission to use this material for any other purposes must be obtained from the IEEE by sending a request to pubs-permissions@ieee.org.};
\end{tikzpicture}

\begin{abstract}
	
%
% Introduction
%
Vehicular big data is anticipated to become the ``new oil'' of the automotive industry which fuels the development of novel crowdsensing-enabled services.
%
% Problem statement
%
However, the tremendous amount of transmitted vehicular sensor data represents a massive challenge for the cellular network.
A promising method for achieving relief which allows to utilize the existing network resources in a more efficient way is the utilization of intelligence on the end-edge-cloud devices. Through machine learning-based identification and exploitation of highly resource efficient data transmission opportunities, the client devices  are able to participate in overall network resource optimization process.
%
% Solution appraoch
%
In this work, we present a novel client-based opportunistic data transmission method for delay-tolerant applications which is based on a hybrid machine learning approach: Supervised learning is applied to forecast the currently achievable data rate which serves as the metric for the reinforcement learning-based data transfer scheduling process. In addition, unsupervised learning is applied to uncover geospatially-dependent uncertainties within the prediction model.
%
% Results
%
In a comprehensive real world evaluation in the public cellular networks of three German \acp{MNO}, we show that the average data rate can be improved by up to $223$~\% while simultaneously reducing the amount of occupied network resources by up to $89$~\%. As a side-effect of preferring more robust network conditions for the data transfer, the transmission-related power consumption is reduced by up to $73$~\%. The price to pay is an increased \ac{AoI} of the sensor data.

\end{abstract}

\maketitle

\section{Introduction} \label{sec:introduction}

The various sensing and communication capabilities of modern vehicles have brought up \emph{vehicular crowdsensing} \cite{Xu/etal/2018a, Ren/etal/2015a} as a novel method for acquiring various kinds of measurement data. Hereby, the mobility behavior of the vehicles is exploited to dynamically cover large areas with sensing capabilities.
It is expected that the \emph{vehicle-as-a-sensor} approach will catalyze the development of data-driven applications such as distributed creation of \ac{HD} environmental maps, traffic monitoring, predictive maintenance, road roughness detection, and distributed weather sensing \cite{Sliwa/etal/2019b}.
%
%
%

%
% Delay-tolerant nature 
%
As pointed out by \cite{Akpakwu/etal/2018a}, a high amount of these target applications --- in particular, \emph{mapping} services --- can be characterized as \emph{delay-tolerant}. Hereby, the applications do not require immediate data delivery but specify soft deadlines within which the received information is considered meaningful. In their empirical analysis, the authors of \cite{Capponi/etal/2019a} analyzed the properties of 32 existing crowdsensing systems from which 23 were found to be compatible with \emph{store-and-forward} data delivery mechanisms.
%
% AECC
%
As an example, the \ac{AECC} has analyzed the requirements for distributed construction of HD environmental maps for automated driving in a recent white paper \cite{AECC/2020a}. For permanent and transient static objects (e.g., road network, surrounding buildings, road work), an update interval in the range of multiple hours is proposed. Even for reporting dynamic obstacles such as other traffic participants, periodic data transfer with an interval of 15~s is considered sufficient.
%
%
%

\begin{figure}[]  	
	\vspace{0cm}
	\centering		  
	\includegraphics[width=1.0\columnwidth]{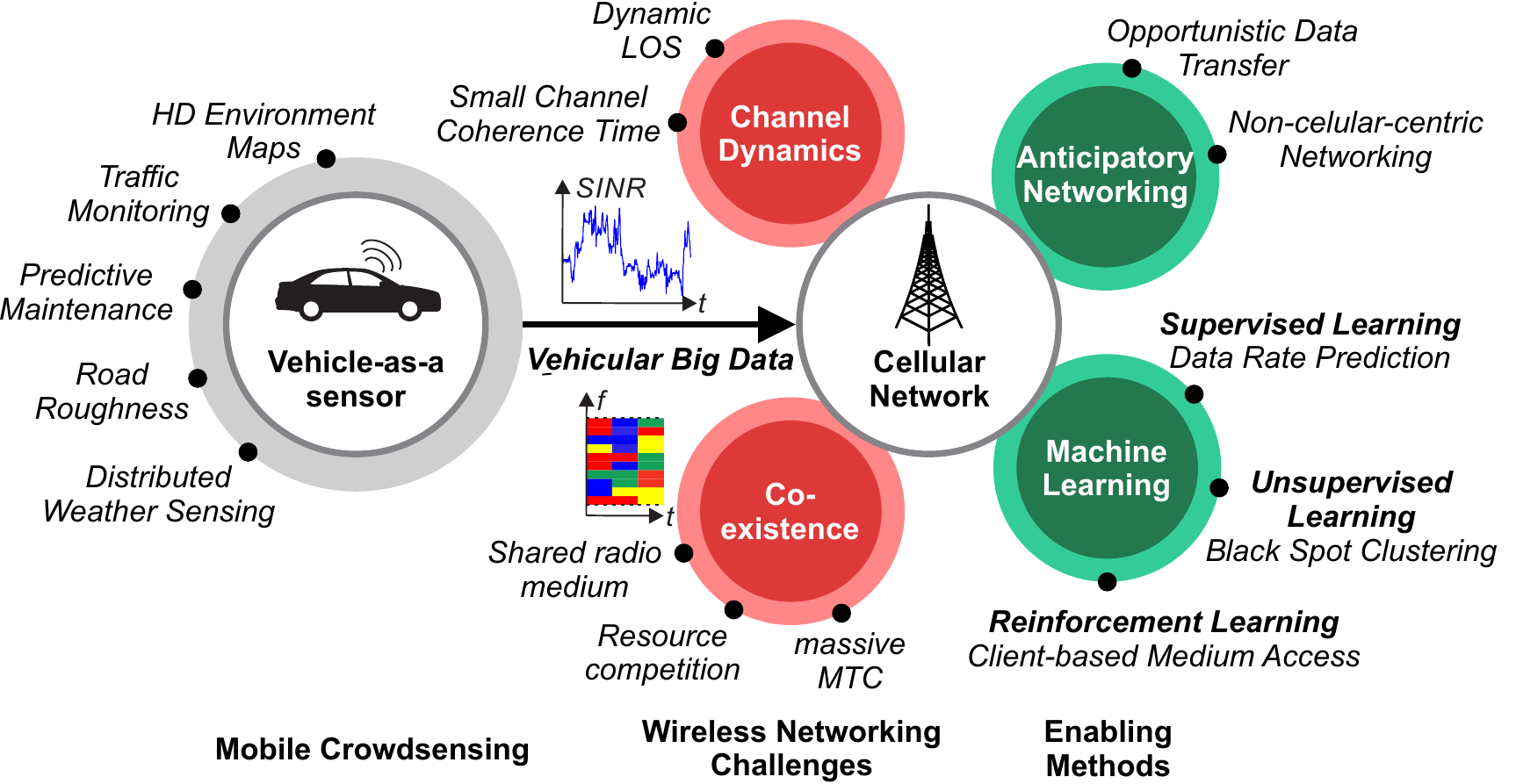}
	\caption{Overview about applications, challenges, and enabling methods for vehicular big data in cellular communication networks.}
	\label{fig:scenario}
	%\vspace{-0.7cm}	
\end{figure}

%
% Problem statement
%
The rise of \emph{vehicular big data} will confront the cellular network with tremendous amounts of resource requirements for vehicular \ac{mMTC}. Since the provision of additional spectrum resources through densification of the network infrastructure is highly cost-intense, it would be preferable to utilize the \emph{existing} resources in a more efficient way through application of machine learning-enabled network intelligence.
An overview about the corresponding applications, challenges, and solution approaches for vehicular big data transfer in cellular networks, which is further described in the following paragraphs, is shown in Fig.~\ref{fig:scenario}. Within the scope of this work, we apply a pragmatic approach which utilizes existing methods from the machine learning domain. However, it is remarked that these enabling methods are themselves subject to active developments in their corresponding research communities. Therefore, it can be expected that future advancements within the neighboring fields can be utilized for further improving the resource efficiency of vehicular big data transfer.
%
%
%

%
% Fig. Scenario
%
%
%
%

%
% 6G: Pervasive intelligence
%
While the current deployments and research efforts for the emerging 5G networks focus on network-side intelligence (e.g., the \ac{NWDAF} allows machine learning-based load analysis of network slices \cite{3GPP/2019a}), researchers agree that \emph{pervasive} intelligence will be one of the key drivers for future 6G networks which are expected to be deployed around 2030 \cite{Yang/etal/2019a, Ali/etal/2020a}.
As a consequence, this will catalyze the development of \emph{non-cellular-centric} networking mechanisms such as \emph{end-edge-cloud} orchestrated intelligence \cite{Ren/etal/2019a} where locally applied machine learning mechanisms allow the client devices to participate in network functions and contribute to the overall network optimization.
%
%
%

\begin{figure}[]  	
	\vspace{0cm}
	\centering		  
	\includegraphics[width=1.0\columnwidth]{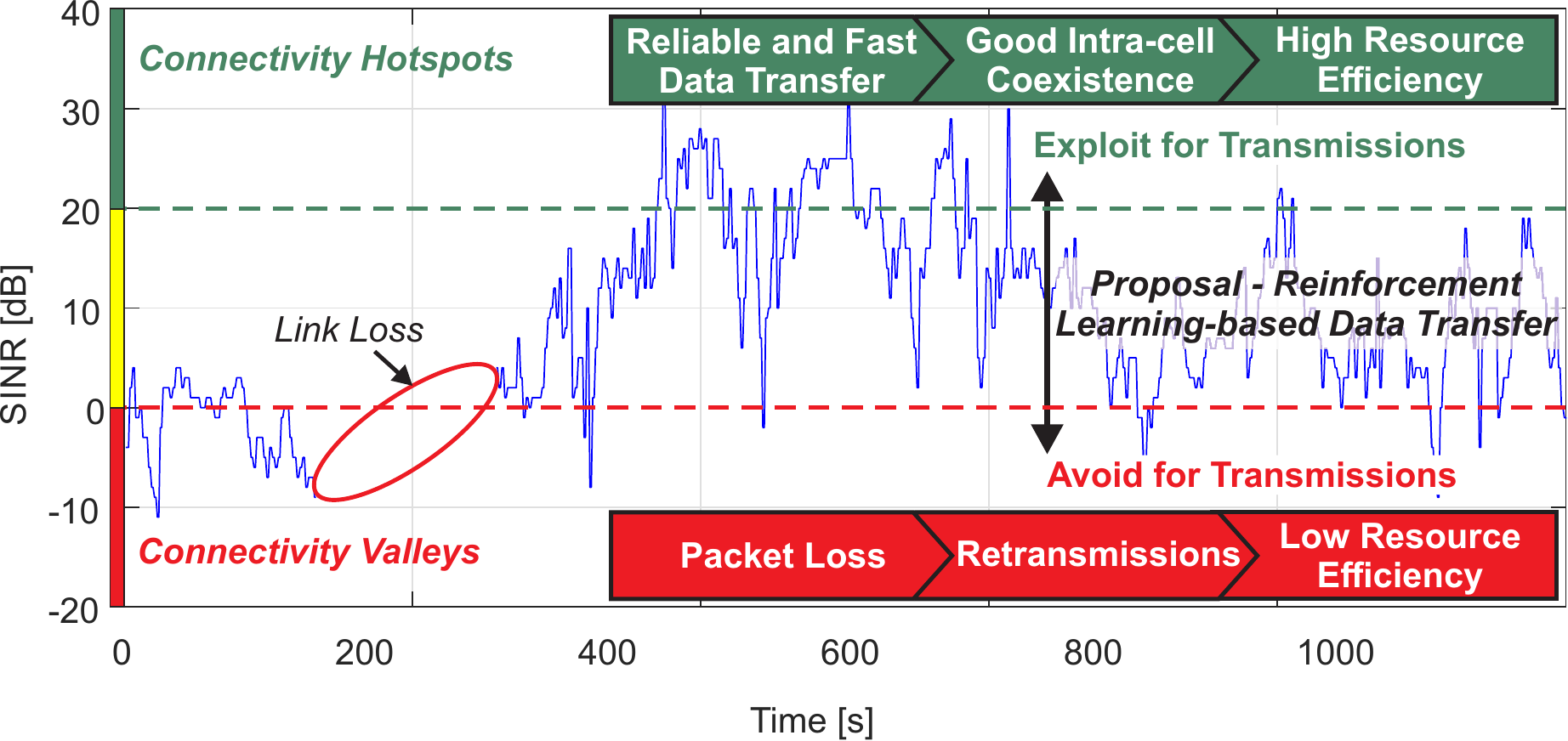}
	\caption{Example for the dynamics of the vehicular radio channel. For optimizing the achievable resource efficiency, client-based intelligence is used to exploit connectivity hotspots and avoid transmissions during connectivity valleys.}
	\label{fig:sinr_trace}
	%\vspace{-0.3cm}	
\end{figure}

%
% Resource wastage
%
An important observation which motivates our contribution is that regular fixed-interval data transmission schemes experience a large variance of the network quality (see Fig.~\ref{fig:sinr_trace}). In order to avoid packet errors and retransmission, the mobile \acp{UE} dynamically adjust the \ac{MCS} to achieve a better robustness in challenging channel situations. However, since lower \acp{MCS} reduce the transmission efficiency and increase the occupation time of the \acp{PRB}, this method results in a wastage of network resources and has a negative impact on the intra-cell coexistence.

%
%  Opportunistic Data Transfer
%
In this work, we exploit the delay-tolerant nature of many vehicular crowdsensing applications as well as the mobility of the vehicles for improving the cellular resource efficiency. Client-based intelligence is applied in order to autonomously schedule the data transfer with respect to the anticipated transmission efficiency. Our proposed method brings together and extends the results of previous work for reinforcement learning-enabled data transfer in vehicular scenarios \cite{Sliwa/Wietfeld/2020a, Sliwa/etal/2020b}.

%
% Contributions
%
The contributions provided by this paper are summarized as follows:
\begin{itemize}
	\item Presentation of \proposal as a novel \textbf{hybrid machine learning} approach for opportunistic data transfer for mobile and vehicular networks.
	\item Comprehensive \textbf{real world} performance analysis and comparison to existing data transfer methods.
	\item Proof-of-concept evaluation for compensating concept drift situations of the data rate prediction through \textbf{online learning}.
	\item The raw results and the developed measurement software are provided in an \textbf{open source} way. \footnote{\url{https://github.com/BenSliwa/rawData_opportunistic_data_transfer}}
\end{itemize}
%
%
%

%
% Fig. SINR trace
%
%
%
%

%
% Structure of the paper
%
The remainder of the paper is structured as follows. After discussing the related work in Sec.~\ref{sec:related_work} and giving an overview about the different evolution stages of the novel method in Sec.~\ref{sec:continuity}, we present the reinforcement learning-based solution approach in Sec.~\ref{sec:rl}. Afterwards, the methodological setup is introduced in Sec.~\ref{sec:methods} and the achieved results are presented and discussed in Sec.~\ref{sec:results}. Based on the resulting insights, we derive recommendations for future 6G networks which are summarized in Sec.~\ref{sec:recommendations}.

\section{Related Work} \label{sec:related_work}

%
% MACHINE LEARNING
%
\textbf{Machine learning} has received tremendous attention within the wireless research community due to its inherent capability of implicitly considering hidden interdepencies between measurable indicators which are too complex to model analytically. Different summary papers \cite{Wang/etal/2020a, Jiang/etal/2017a, Ye/etal/2018a, Sun/etal/2019a} provide comprehensive information about using machine learning methods for optimizing wireless networks. Three major machine learning disciplines are distinguished:
\begin{itemize}
	%
	% Supervised learning
	%
	\item \textbf{Supervised} learning allows to learn a model $f_{\text{ML}}$ on \emph{features} $\mathbf{X}$ with \emph{labeled} data $\mathbf{Y}$ such that $f: \mathbf{X} \rightarrow \mathbf{Y}$. After the training phase, the model can be utilized to make predictions $\tilde{y}$ on novel unlabeled data $\mathbf{x}$ such that $\tilde{y} = f(\mathbf{x})$.
	%
	% Regression models
	%
	For this purpose, popular model classes are (deep) \acp{ANN} \cite{LeCun/etal/2015a}, \acp{CART}-based methods such as \acp{RF} \cite{Breiman/2001a}, and Bayesian models such as \ac{GPR} \cite{Rasmussen/2004a}.

	%
	% Unsupervised models
	%
	\item \textbf{Unsupervised} learning is applied to cluster measurements based on patterns in non-labeled data sets. A popular method for this category is the k-means \cite{Arthur/Vassilvitskii/2007a} algorithm.

	%
	% Reinforcement learning
	%
	\item \textbf{Reinforcement} learning \cite{Gacanin/2019a, Sutton/Barto/2018a} teaches an \emph{agent} to autonomously perform favorable \emph{actions} in a defined \emph{environment} by learning from the observed \emph{rewards} of previously taken actions.
	%
	% Q-Learning
	%
	Q-Learning \cite{Watkins/Dayan/1992a} represents the foundation for most more complex methods such as deep reinforcement learning.
\end{itemize}
%
%
%

%
% Fig. CAT Continuity
%
\begin{figure*}[t]  	
	\centering		  
	\includegraphics[width=1.0\textwidth]{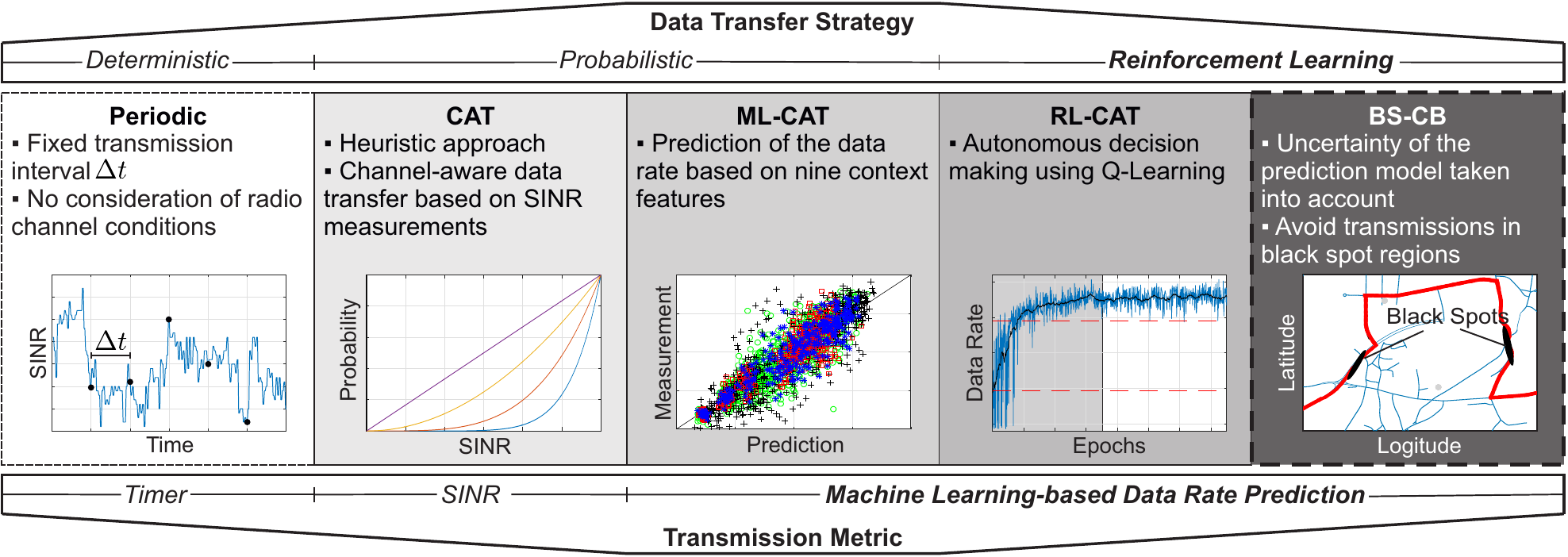}
	\caption{Continuity of context-aware approaches for opportunistic data transmissions in vehicular networks.}
	\label{fig:continuity}	
	%\vspace{-0.7cm}
\end{figure*}
%
%
%

%
% 3GPP Initiatives
%
Within commercial deployments of emerging 5G networks, the implementation of machine learning-based intelligence mainly focuses on the network infrastructure side. \ac{NWDAF} \cite{3GPP/2019a, Sevgican/etal/2020a} is a novel machine learning-enabled network function which is used by the \acp{MNO} to determine and predict the network load. Different use-cases that could exploit this information --- e.g., traffic routing, mobility management, load balancing, and handover optimization --- are motivated in \cite{3GPP/2019b}.
%
% Ali/etal/2020a
%
Among others, the white paper of \cite{Ali/etal/2020a} envisions \emph{pervasive machine learning} as one of the fundamental enabling methods for future 6G networks which are expected to be deployed around 2030.
%
% Park/etal/2019a: ML at the edge
%
As a consequence of the trend of bringing intelligence closer towards the client devices, resource-aware machine learning has become an emerging research topic. A comprehensive summary about resource aspects for edge-based intelligence is provided by Park et al. in \cite{Park/etal/2019a}.

%
% End-to-end
%
The recent advancements in machine learning-based data analysis have also led to the rise of the end-to-end modeling paradigm for wireless communication systems \cite{Doerner/etal/2018a} and have catalyzed the development of novel data-driven performance evaluation methods.
%
% DDNS
%
\ac{DDNS} \cite{Sliwa/Wietfeld/2019d, Sliwa/Wietfeld/2019c} allows to analyze the performance of wireless communication systems by replaying empirically acquired context traces. The end-to-end behavior of the observed target \ac{KPI} is then derived by a combination of deterministic and probabilistic machine learning models which mimics the statistical derivations of the real world measurements. In comparison to conventional system-level network simulation \cite{Cavalcanti/etal/2018a}, this method is able to achieve a better modeling accuracy of radio propagation effects in concrete real world evaluation scenarios and achieves a massively higher computational efficiency. Another advantage is a reduction of the simulation setup complexity since the end-to-end analysis approach solely relies on the acquired data and does not require to parameterize communicating entities.

%
% ANTICIPATORY COMMUNICATION
%
\textbf{Anticipatory mobile networking} \cite{Bui/etal/2017a} is a novel wireless communications paradigm which aims to optimize decision processes in communication systems through explicit consideration of \emph{context} information. Since mobile and vehicular networks are inherently impacted by the interdependency of mobility and radio channel dynamics \cite{Toufga/etal/2019a}, machine learning-enabled anticipatory networking is a promising approach for system optimization in this domain. As an example, Dalgkitsis et al. \cite{Dalgkitsis/etal/2020a} utilize mobility prediction jointly with deep learning for improving the service orchestration process in 5G vehicular networks.

%
% Coll-Perales/etal/2019a
%
\emph{Non-cellular-centric} networking \cite{Coll-Perales/etal/2019a} integrates the network clients as part of the network fabric and allows them to contribute explicitly or implicitly to network management functions.
%
% Delay-tolerant applications
%
This approach allows to exploit the capability of the clients to sense their environments for opportunistically scheduling data transmissions for delay-tolerant applications \cite{Ha/etal/2012a} in a context-aware manner.
%
% Shi/etal/2014a : CoAST
%
In \cite{Shi/etal/2014a}, Shi et al. point out that network congestion has a large short-term variance and that traffic peaks can be compensated by delaying transmissions. Therefore, the authors propose the \ac{CoAST} system which applies a collaborative infrastructure-assisted optimization approach based on dynamic pricing. Hereby, the announced traffic demands of the \acp{UE} are used by a central entity which computes and broadcasts the projected data transfer prices for a given future time window. This information is then used by the \acp{UE} to schedule their transmissions with respect to the trade-off between price and additional delay.
%
% Peek-n-sneak & CASTLE
%
Peek-n-sneak \cite{Chakraborty/etal/2013a} and \ac{CASTLE} \cite{Lee/etal/2019a} are distributed transmission scheduling approaches which rely on a threshold decision for performing or delaying the data transfer. Both approaches use different network quality indicators (\ac{RSRP}, \ac{RSRQ}, and \ac{SINR}) for predicting the current network load based on a \ac{RBF} \ac{SVM}.

%
% DATA RATE PREDICTION
%
\textbf{Data rate prediction} can serve as a metric for anticipatory decision making such as opportunistic data transfer \cite{Sliwa/etal/2019d} and dynamic \ac{RAT} selection. The predictions can either be performed \emph{actively} or \emph{passively}.
%
% Active vs passive
%
Active prediction methods apply time series analysis -- e.g., based on \ac{LSTM} methods as considered in \cite{Nikolov/etal/2020a, Lee/etal/2020a} -- and monitor the behavior of ongoing data transmissions. Since the need to continuously transmit data is opposed to the considered opportunistic medium access strategy, this work focuses on passive prediction approaches which have been investigated by different authors. The key insights are summarized as follows:
%
% Key findings for passive data rate prediction
%
\begin{itemize}
	%
	% Radio channel indicators
	%
	\item Radio channel indicators (e.g., defined according to 3GPP TS 36.213 \cite{3GPP/2018b}) are highly correlated to the observed data rate and can serve as meaningful information for predicting the latter \cite{Samba/etal/2017a, Herrera-Garcia/etal/2019a, Riihijarvi/Mahonen/2018a}.
	%
	% Deep learning vs simpler models
	%
	\item Due to the \emph{curse of dimensionality} \cite{Zappone/etal/2020a}, complex models such as \ac{ANN}-based \emph{deep learning} approaches require a significantly higher amount of training data than simpler methods such as \acp{CART}. As typical data sets in the wireless communication domain are comparably small \cite{Ali/etal/2020a}, less complex methods often achieve a higher prediction accuracy \cite{Jomrich/etal/2018a, Sliwa/Wietfeld/2019c}.
	%
	% Payload size
	%
	\item For the derivation of generalizable prediction models, it is important to integrate application-layer knowledge about the \emph{payload size} of the data packet to be transmitted \cite{Sliwa/Wietfeld/2019b}. This way, the prediction is able to implicitly account for the interdependency between transmission duration and channel coherence time as well as payload-overhead-ratio and protocol-specific aspects such as the slow start mechanism of the \ac{TCP}.
	%
	% Model granularity
	%
	\item A low data aggregation granularity should be preferred: Few models with large data sets (e.g., a single prediction model per \ac{MNO}) achieve a better average prediction performance than a large amount of highly-specific models (e.g., dedicated prediction models for each \ac{eNB}) \cite{Sliwa/Wietfeld/2019b, Domingos/2012a}.
	%
	% Time of day
	%
	\item Although temporal effects have a significant impact on the network load, the time of day is negligible if load-dependent network quality indicators such as \ac{RSRQ} are considered in the data set \cite{Akselrod/etal/2017a, Sliwa/Wietfeld/2019b}.
\end{itemize}
%
%  Sliwa/etal/2020a: Cooperative prediction
%
In addition to these purely client-based approaches, the authors of \cite{Sliwa/etal/2020a} have analyzed a possible implementation for \emph{cooperative} data rate prediction in future 6G networks where the network infrastructure actively announces network load information to the mobile clients. In an initial feasibility study, it is shown that the cooperative approach is able to reduce the \ac{RMSE} by 25~\% in uplink and 30~\% in downlink direction

\section{Towards Reinforcement Learning-enabled Opportunistic Data Transfer} \label{sec:continuity}

Different opportunistic data transfer methods have build the foundation for the proposed \proposal method. The different evolution stages are shown in Fig.~\ref{fig:continuity}.

%
% Periodic
%
\textbf{Periodic} data transfer represents the regular approach for transmitting \ac{MTC} data. The medium access is based on a fixed timer interval $\Delta t$ which transmits the data regardless of the radio channel conditions. 

%
% CAT
%
\textbf{\acf{CAT}} \cite{Ide/etal/2015a} is a probabilistic opportunistic data transfers method which schedules the medium access based on measurements of the \ac{SINR}. Data is buffered locally until a transmission decision is made for the whole buffer. The transmission probability $p_{\text{TX}}(t)$ is computed as
%
% Eq. CAT
%
\begin{equation} \label{eq:cat_formula}
	p_{\text{TX}}(t) = \begin{cases}
	0 & \Delta t<\Delta t_{\min} \\ 
	1 & \Delta t>\Delta t_{\max} \\ 
	\left( \frac{\Phi(t)-\Phi_{\min}}{\Phi_{\max}-\Phi_{\min}} \right)^{\alpha} & \text{else}
	\end{cases}
\end{equation}
with $\Phi$ being the transmission metric -- the $\text{SINR}(t)$ measurement -- with a defined value range $\left\lbrace \Phi_{\min}, \Phi_{\max} \right\rbrace$.
$\Delta t$ represents the time since the last transmission has been performed. $\Delta t_{\min}$ is used to guarantee a minimum packet size and  $\Delta t_{\max}$ ensures that the \ac{AoI} does not exceed the requirements of the target application. The exponent $\alpha$ allows to control the preference of high metric values within the data transfer process.

%
% ML-CAT
%
\textbf{\acf{ML-CAT}} \cite{Sliwa/etal/2018b, Sliwa/etal/2019d} is a machine learning-based extension to \ac{CAT}. Due to the short-term fluctuations of the \ac{SINR}, the transmission decision is performed based on data rate predictions which are obtained from an \ac{RF} model (see Sec.~\ref{sec:prediction}). While the actual transmission is still performed based on Eq.~\ref{eq:cat_formula}, the considered metric $\Phi$ corresponds to the predicted data rate $\tilde{S}(t)$.

%
% RL-CAT
%
\textbf{\acf{RL-CAT}} \cite{Sliwa/Wietfeld/2020a} is a first reinforcement learning-based variant of the \ac{ML-CAT} method which replaces the probabilistic medium access with a Q-learning approach aiming to maximize the data rate of the \emph{individual} sensor data transmissions. The predicted data rate and the elapsed buffering time form the context tuple $\mathbf{c_{t}} = (\tilde{S}(t), \Delta t)$ are used to lookup up the action --- \texttt{IDLE} or \texttt{TX} --- with the highest Q-value from a Q-table. The latter is trained as
\begin{equation} \label{eq:q_learning}
	\q = (1-\alpha) \cdot \q + \alpha \left[ r_{a} + \lambda  \cdot \max_a \qq \right] 
\end{equation}
whereas $\alpha$ corresponds to the learning rate, $r_{a}$ is the reward of the action $a$, $\lambda$ represents the discount factor, and $\mathbf{c_{t+1}}$ is an estimation for the $Q$-value after $a$ has been executed.
In classical Q-Learning, it is assumed that the decision making of the agent causes a sequential improvement of its \emph{state} within the environment and ultimately leads to reaching an ``optimal'' target state. However, as further discussed Sec.~\ref{sec:rl}, in the considered opportunistic data transfer use case, the agent-related impact on the state of the environment is negligible due to the dominance of external influences such as the channel and network load dynamics: Even if the agent was capable of performing hypothetical ``optimal'' actions, its state --- represented by the context tuple  $\mathbf{c_{t}}$ --- would be still determined by the impact of the non-controllable influence factors. Therefore $\lambda$ is set to $0$ which results in a simplified Q-Learning variant 
%
% Eq. Proposed Q update
%
\begin{equation}
	\q = (1-\alpha) \cdot \q + \alpha \cdot r_{a} .
\end{equation}	
that implements a myoptic approach focusing on optimizing the immediate reward of the taken actions.

\section{Reinforcement Learning-based Opportunistic Data Transfer with BS-CB} \label{sec:rl}

In this section, we present the novel \proposal method.
%
% Zhou/etal/2019a: Edge intelligence
%
According to the classification scheme for edge intelligence provided by \cite{Zhou/etal/2019a}, the proposed data transfer method represents a  \emph{level 3: on-device inference} edge intelligence implementation where the model is trained in the cloud/offline and inference is run completely locally.

%
% Fig. Reinforcement learning
%
\begin{figure}[]  	
	\vspace{0cm}
	\centering		  
	\includegraphics[width=1.0\columnwidth]{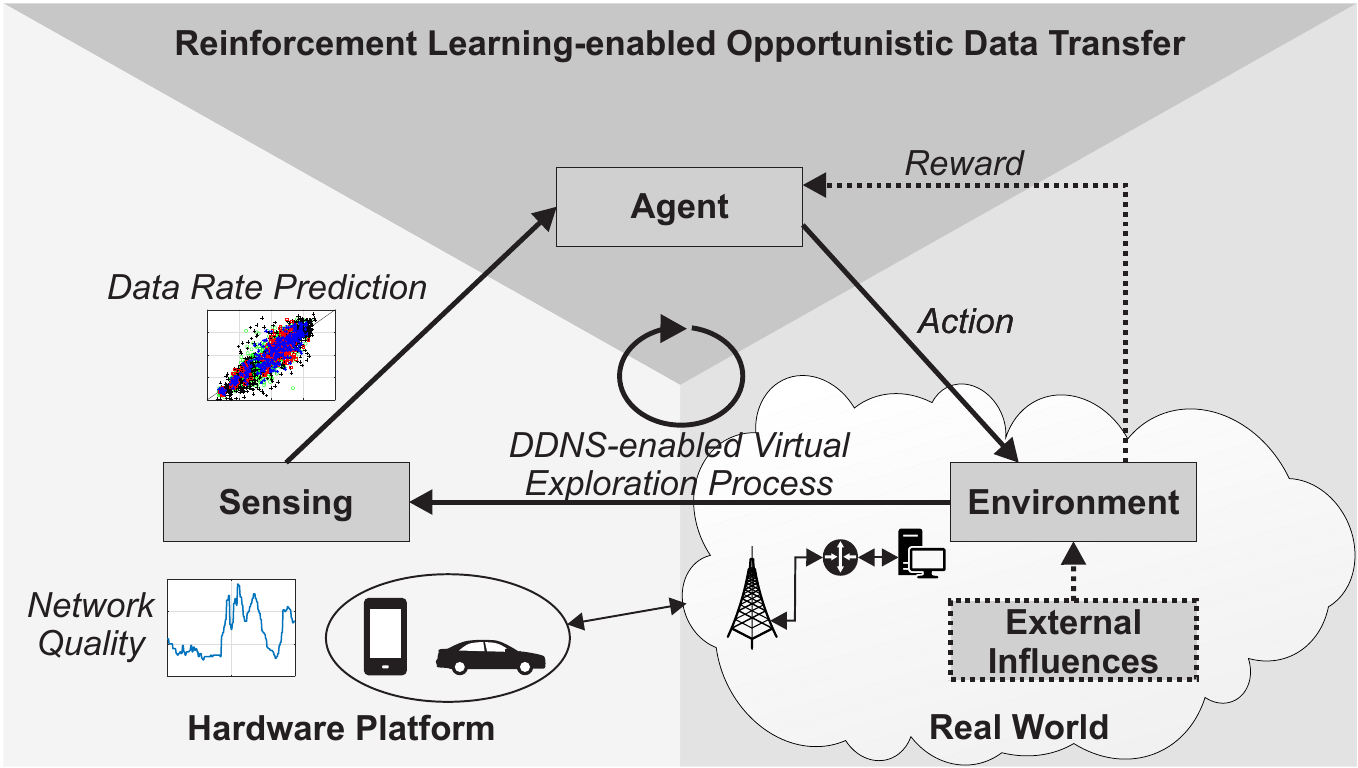}
	\caption{Interaction between the different logical entities within a reinforcement learning setup for opportunistic data transfer.}
	\label{fig:reinforcementLearning}
	%\vspace{0cm}	
\end{figure}
A schematic overview about the interaction between the different logical entities is shown in Fig.~\ref{fig:reinforcementLearning}. 
%
% Reinforcement learning
%
\begin{itemize}
	%
	% Agent
	%
	\item The actual opportunistic data transfer is modeled as a reinforcement learning \textbf{agent} which senses its \emph{environment}, performs \emph{actions} and observes the resulting \emph{rewards}.
	%
	% Environment
	%
	\item Hereby, the \textbf{environment} is represented by the real world cellular network. Classical reinforcement learning assumes that the actions taken by the agent change the \emph{state} of the environment. However, in the considered vehicular scenarios, the properties of the environment are highly \emph{time-variant} due to the dynamically changing radio channel conditions mainly related to the mobility behavior of the mobile \ac{UE}. In addition, other users of the cellular network consume network resources which leads to the conclusion that the state of the environment mainly depends on the \emph{external influences}.
	%
	% Sensing
	%
	\item The \textbf{sensing} of the environment is performed through the hardware platform which observes context indicators. In order to reduce the dimensionality of the reinforcement learning problem, data rate prediction is applied.
\end{itemize}
%
%
%

%
% Fig. Architecture
%
\begin{figure}[]  	
	\vspace{0cm}
	\centering		  
	\includegraphics[width=1.0\columnwidth]{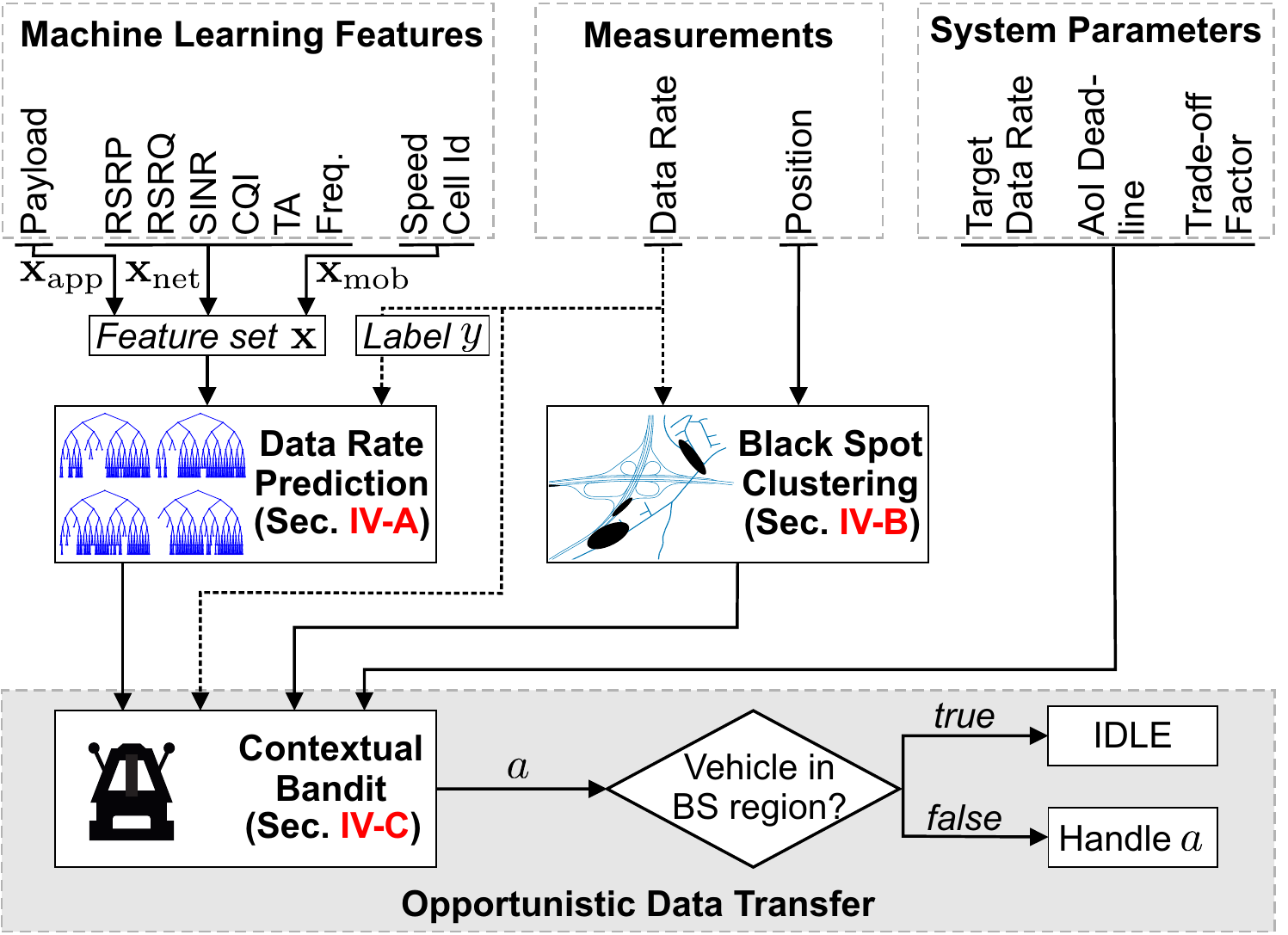}
	\caption{Overall system architecture model of the proposed \proposal method.}
	\label{fig:cb_architecture}
	%\vspace{-0.7cm}	
\end{figure}
The overall system model of the novel \proposal is shown in Fig.~\ref{fig:cb_architecture}. \proposal implements a hybrid approach which brings together all major machine learning disciplines. Supervised learning is applied to predict the achievable data rate based on measured context indicators. Unsupervised learning is then utilized to detect geospatially-dependent uncertainties of the prediction model. Finally, the reinforcement learning-based autonomous data transfer uses the acquired information for optimizing the resource efficiency of vehicular data transmissions.
In the following paragraphs, the three main components of the proposed methods are introduced in further details.

\subsection{Supervised Learning: Data Rate Prediction} \label{sec:prediction}

The overall feature set $\mathbf{x}$ is composed of nine different features from multiple context domains:
%
% Context domains
%
\begin{itemize}
	\item \textbf{Network context} $\mathbf{x}_{\text{net}}$: \ac{RSRP}, \ac{RSRQ}, \ac{SINR}, \ac{CQI}, \ac{TA}
	\item \textbf{Mobility context} $\mathbf{x}_{\text{mob}}$: Velocity of the vehicle, cell id of the connected \ac{eNB}
	\item \textbf{Application context} $\mathbf{x}_{\text{app}}$: Payload size of the data packet to be transmitted
\end{itemize}
The data rate is then predicted based on a regression model $f_{\text{ML}}$ as $\tilde{S}(t) = f_{\text{ML}}(\mathbf{x})$. As a preprocessing step, we compare the prediction performance of different machine learning models whereas the parameterization of each model has been optimized based on grid search:
%
% Prediction models
%
\begin{itemize}
	%
	% ANN
	%
	\item \textbf{\acf{ANN}}  \cite{LeCun/etal/2015a} with two hidden layers with 10 neurons per hidden layers and sigmoid activation function, momentum $\alpha=0.001$, learning rate $\eta=0.1$, and 500 training epochs. 
	%
	% M5 + RF
	%
	\item \ac{CART} methods \textbf{\acf{M5}} and \textbf{\acf{RF}} \cite{Breiman/2001a} with 100 random trees and maximum depth 15.
	%
	% SVM
	%
	\item \textbf{\acf{SVM}} with \ac{RBF} kernel and \ac{SMO} training.
\end{itemize}
%
%
%

%
% Fig. RMSE
%
\begin{figure}[]  	
	\vspace{0cm}
	\centering		  
	\includegraphics[width=1.0\columnwidth]{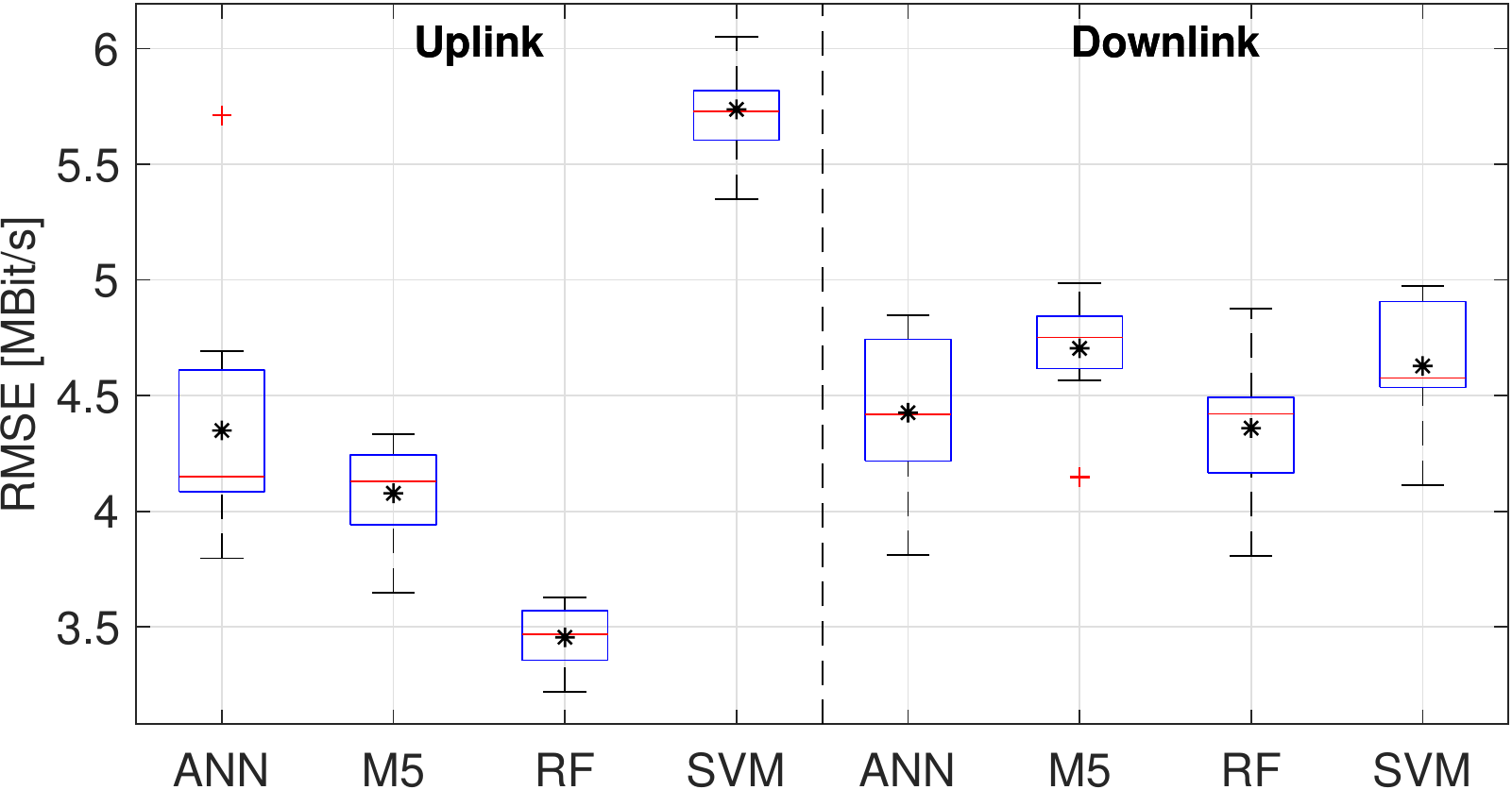}
	\caption{Resulting data rate prediction performance for different regression models on the \mno{A} data set. \emph{ANN:} Artificial Neural Network, \emph{M5}: M5 Regression Tree, \emph{RF}: Random Forest, \emph{SVM}: Support Vector Machine}
	\label{fig:rmse}
	%\vspace{-0.5cm}	
\end{figure}
The resulting \ac{RMSE} of the data rate prediction models on the \mno{A} data set of \cite{Sliwa/Wietfeld/2019b} is shown Fig.~\ref{fig:rmse}. In both evaluations, the lowest prediction error is achieved by the \ac{RF} model.
In uplink direction, different context indicators have specific regions of application: As discussed in \cite{Sliwa/Wietfeld/2019b}, \ac{RSRQ} is an important indicator for the data rate in cell edge regions and \ac{SINR} has a higher impact on the latter in the center of the cell --- both can be distinguished through considering the \ac{RSRP}. These interval-wise scope regions match well with the condition-based model architecture of the \ac{RF} model.
%
% Uplink vs Downlink
%
However, in downlink transmission direction, the differences between the considered prediction models are less significant. This observation can be explained through consideration of the findings of \cite{Bui/etal/2017a}: In downlink direction, the resulting data rate is mostly related to the cell load which is partially represented by the \ac{RSRQ}. The presence of this dominant feature results in a less complex learning task. Since the \ac{RSRQ} is only an implicit indicator for the current network load, the resulting \ac{RMSE} is relatively high.

%
% Focus on RF
%
Due to these observations, we apply the \ac{RF} model for performing the context-based data rate predictions in the remainder of this paper.

\subsection{Unsupervised Learning: Black Spot Clustering}

%
% Motivation
%
An important observation of previous work \cite{Sliwa/Wietfeld/2020a} is that the resulting data rate prediction accuracy in vehicular scenarios has a \emph{geospatial dependency}: Large outliers often occur \emph{cluster-wise} due to local effects such as \ac{eNB} handovers, cell switches, and environment-dependent sporadic link loss.
Although the knowledge about these mechanisms does not explicitly allow us to compensate the undesired effects, it can be exploited within the opportunistic data transmission processes as a measurement for the \emph{uncertainty} of the prediction model: Transmissions should be avoided if the prediction model is currently in an unreliable state and does not allow to make a precise statement about the achievable end-to-end performance.
%
% Black spot definition
%
We call these areas \emph{black spots} based on the usage of the term in traffic satefy where it refers to regions with a significantly increased probability for collisions of vehicles.

\begin{figure}[]  	
	\vspace{0cm}
	\centering		  
	\includegraphics[width=1.0\columnwidth]{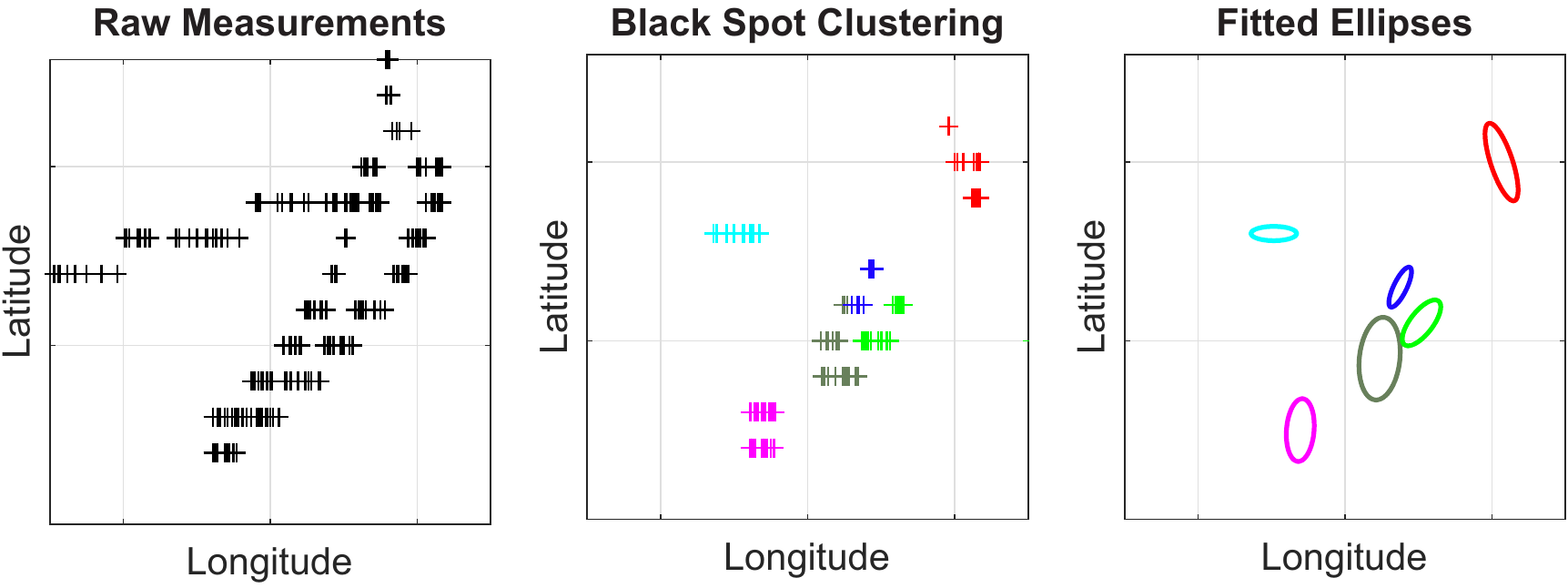}
	\caption{Steps of the black spot clustering process.}
	\label{fig:clustering}
	%\vspace{-0.5cm}	
\end{figure}

%
% BLACK SPOT
%
The proposed black spot-aware networking approach is divided into the unsupervised learning-based offline data analysis and the online application.
%
% OFFLINE DATA ANALYSIS
%
The \textbf{offline data analysis} consists of multiple steps which are visualized in Fig.~\ref{fig:clustering}.
%
% Fig. Clustering
%
%
% Clustering steps
%
\begin{enumerate}
	%
	% Geo-clustering
	%
	\item \textbf{Geo-clustering}: Unsupervised learning based on \emph{k-means} \cite{Arthur/Vassilvitskii/2007a} is applied in order to cluster the transmission locations into a total amount of $N_{c}$ clusters. 
	%
	% Black spot detection
	%
	\item \textbf{Black spot detection}: For each cluster $c$, the \ac{RMSE} (see Sec.~\ref{sec:methods}) of the data rate prediction results is computed and compared to a threshold value $\text{RMSE}_{\text{max}}$. All clusters that exceed the given upper limit are labeled as \emph{black spot clusters}.
	%
	% Ellipse fitting
	%
	\item \textbf{Ellipse fitting}: All detected black spot clusters are fitted to rotated ellipses in order to allow their later online consideration within the opportunistic data transmission process. Hereby, the length $a$ of the ellipse is calculated based on the dominant intra-cluster distance vector. 
\end{enumerate}
%
%
%

%
% Fig. GPR
%
\renewcommand{\boxWidth}{0.32\textwidth}
\begin{figure*}[] 
	\centering
	\subfloat[Overall prediction model]{\includegraphics[width=\boxWidth]{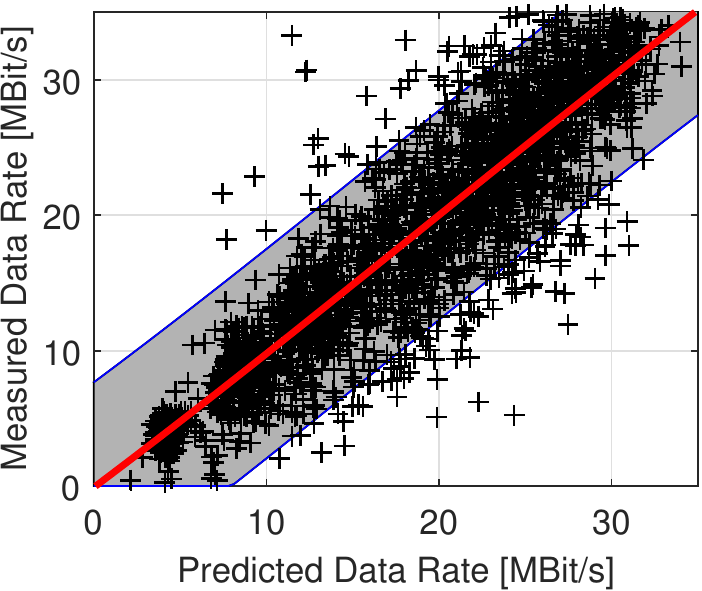}}\hfill
	\subfloat[Non-black spot prediction model]{\includegraphics[width=\boxWidth]{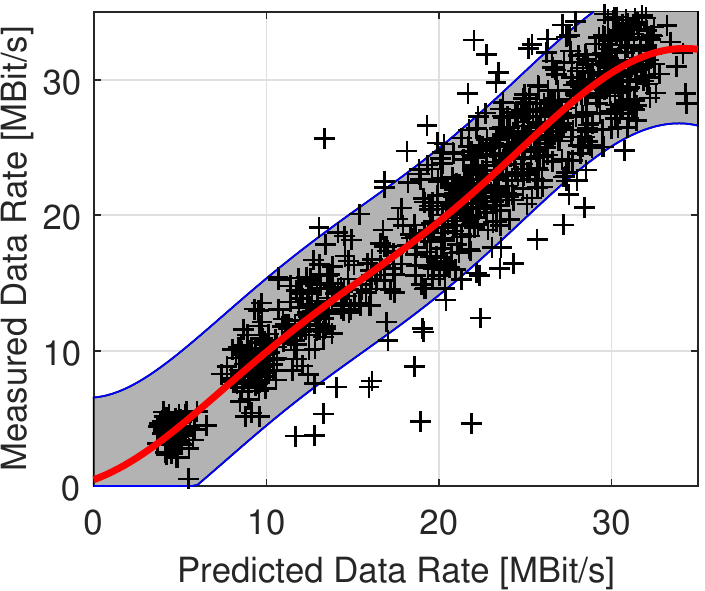}}\hfill	
	\subfloat[Only black spot prediction model]{\includegraphics[width=\boxWidth]{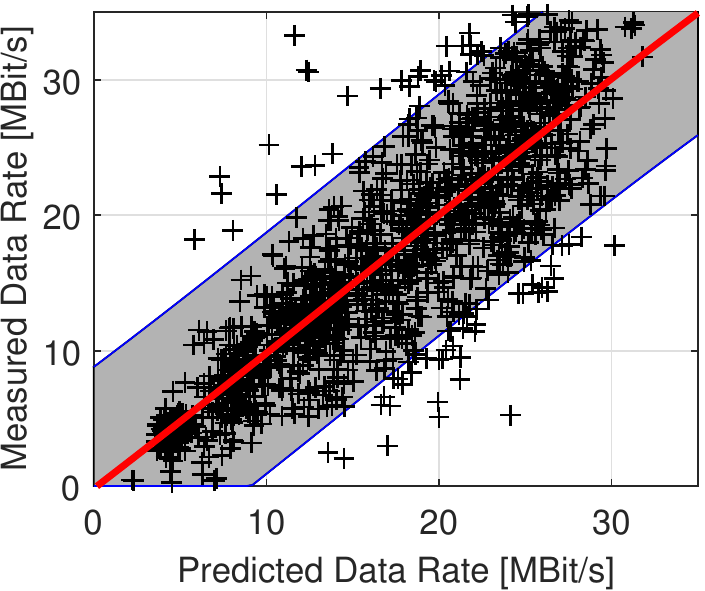}}\hfill		
	\caption{The overall prediction model is separated into a more precise model for non-black spot regions and a less precise model for black spot regions. The gray area shows the behavior of a 0.95-confidence area derived by applying a \ac{GPR} model on the results of the prediction model.}
	\label{fig:gpr}
	%\vspace{-0.5cm}
\end{figure*}
The impact of considering information about black spot regions within the prediction model is shown in Fig.~\ref{fig:gpr}. While Fig.~\ref{fig:gpr}~(a) shows the resulting prediction performance of the overall data set which consists of black spot and non-black regions, the separation of the prediction model allows to improve the prediction accuracy for the non-black spot regions as shown in Fig.~\ref{fig:gpr}~(b). In the following, we will use this variant for predicting the data rate as the metric of the opportunistic data transmission process.

\begin{figure}[]  	
	\vspace{0cm}
	\centering		  
	\includegraphics[width=1.0\columnwidth]{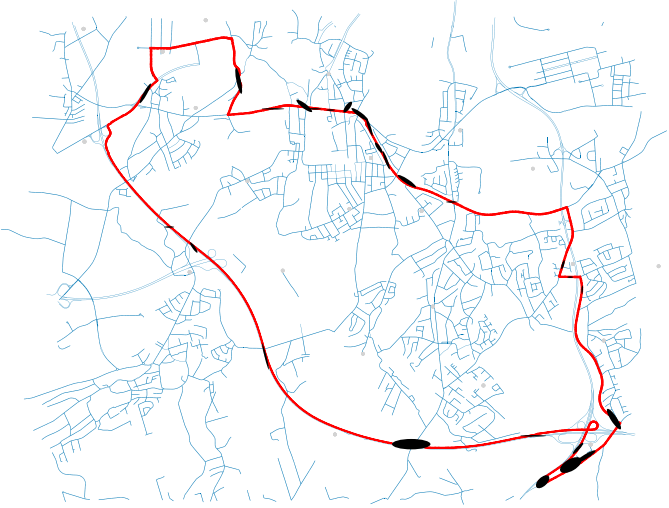}
	\caption{Resulting black spot regions along the evaluation track for \mno{A} in uplink direction (Map: ©OpenStreetMap contributors, CC BY-SA).}
	\label{fig:map}
	%\vspace{-0.7cm}	
\end{figure}

%
% ONLINE APPLIICATION
%
For the \textbf{online application}, the vehicle's position $\mathbf{P}$ is compared against all black spot ellipses with corresponding ellipse centroid $\mathbf{P}_i$ based on an intersection test for $\alpha$-rotated ellipses. The vehicle is within the considered elliptic region if the following condition is fulfilled:
%
% Eq. Point in Ellipse (with rotation)
%
\begin{equation}
	\frac{(c \cdot \mathbf{v}.x + s \cdot \mathbf{v}.y)^2}{a^2}
	+ \frac{(s \cdot \mathbf{v}.x - c \cdot \mathbf{v}.y)^2}{b^2}
	\leq 1
\end{equation}
with $\mathbf{v} = \mathbf{P} - \mathbf{P}_i$, $c = \cos \alpha$, $s = \sin \alpha$, and $\alpha$ being the ellipse rotation. An overview about the detected black spot regions for \mno{A} in uplink direction is shown in Fig.~\ref{fig:map}.

%
% Fig. Black spot map
%
%
%
%

%
% CONTEXTUAL BANDIT
%
\subsection{Reinforcement Learning: Contextual Bandit-based Data Transfer}

%
% CONTEXTUAL BANDIT
%
The actual opportunistic data transfer is modeled as a \ac{LinUCB} \cite{Li/etal/2010a} contextual bandit whereas the \emph{arms} of the bandit correspond to the possible actions:
%
% Actions
%
\begin{itemize}
	%
	% IDLE
	%
	\item $\mathbf{a}_{\textbf{IDLE}}$ leads to a local buffering of the newly acquired data as the current network quality is not considered appropriate for allowing resource efficient data transfer. It is assumed that due the mobility behavior of the vehicle, the mobile \ac{UE} will encounter a more suitable transmission opportunity in the future.
	%
	% TX
	%
	\item $\mathbf{a}_{\textbf{TX}}$ causes the transmission of the whole buffered data.
\end{itemize}

The context-aware arm selection process is performed based on a sequence of matrix-vector multiplications as
%
% Eq. Arm selection
%
\begin{equation}
	a = \argmax_{a\in \mathbf{A}} \left(
	\underbrace{\hat{\theta}^{T}_{a} \mathbf{c}}_{\text{Estimated reward}} + 
	\underbrace{\alpha \sqrt{\mathbf{c}^{T} \mathbf{A}^{-1}_{a} \mathbf{c}}}_{\text{UCB}}
	\right) .
\end{equation}
%
% Estimated reward
%
Hereby, the estimated reward is derived by ridge regression whereas $\hat{\theta}_{a}$ represents the regression coefficients of arm $a$ which are updated during the reinforcement learning process and $\mathbf{c} = (\tilde{S}(t), \Delta t)$ is the $d$-dimensional context tuple consisting of the predicted data rate $\tilde{S}(t)$ and the current buffering time $\Delta t$.
%
%
%

%
% UCB Part
%
$\mathbf{A}_{a}$ is computed as $\mathbf{A}_{a} = \mathbf{D}^{T}_{a} \mathbf{D}_{a} + \mathbf{I}_{a}$
with $\mathbf{I}_{a}$ being a $d$-dimensional identity matrix and $\mathbf{D}_{a}$ being the $m \times d$ matrix which contains the $m$ previously observed context tuples.
The constant exploration parameter $\alpha$ controls the greediness of the algorithm and is computed as 
%
% Eq. Alpha
%
\begin{equation}
	\alpha =  1 + \sqrt{\frac{\ln(2/\delta))}{2}}
\end{equation}
based on the only system parameter $\delta$. The smaller the value of $\alpha$, the more greedy the algorithm behaves, meaning that it will more likely \emph{exploit} actions that currently seem to be optimal. 
After each performed action, the regression coefficients are updated based on the observed reward $r_{a}$ as
%
% Eq. Regression coefficients
%
\begin{equation}
	\hat{\theta}_{a} \leftarrow \mathbf{A}^{-1}_{a} \mathbf{b}_{a}
\end{equation}
with
%
% Eq. 
%
\begin{equation}
	\mathbf{b}_{a} \leftarrow \mathbf{b}_{a} + r_{a} \cdot \mathbf{c}
\end{equation}
Hereby, $\mathbf{b}_{a}$ is initialized as a $d$-dimensional zero vector. The reward functions are computed action-specific, for the \texttt{TX} action, the reward is derived as:
%
% Reward functions
%
%
% Eq. TX Reward
%
\begin{equation} \label{eq:tx_reward}
	r_{\text{TX}}(S, \Delta t) = \frac{\omega \cdot (\tilde{S}-S^{*})}{S_{\max}} + \frac{\Delta t \cdot (1-\omega)}{\Delta t_{\max}} 
\end{equation}
whereas the trade-off factor $w$ controls the fundamental trade-off between data rate optimization and \ac{AoI} optimization. $S^{*}$ represents a target data rate which should be approached and $S_{\max}$ is the empirically observed maximum data rate of the network. $\Delta t$ is an application-specific deadline for the tolerable \ac{AoI}.

The reward of the \texttt{IDLE} action is computed as:
%
% Eq. IDLE Reward
%
\begin{equation} \label{eq:idle_reward}
	r_{\text{IDLE}}(\Delta t) = \begin{cases}
	\Omega & \Delta t \geq \Delta t_{\max} \\ 
	0 & \text{else}
	\end{cases}
\end{equation}
whereas $\Omega$ is chosen as a negative number which ensures that the estimated reward of the \texttt{TX} action is superior to the reward of the \texttt{IDLE} action if $\Delta t$ exceeds the \ac{AoI} deadline $\Delta t_{\max}$. As a result, the data is transferred immediately regardless of the radio channel conditions.
%
%
%

%
% No transmissions within black spot regions
%
After the contextual bandit has made a transmission decision, the information about the black spot regions is leveraged: If the vehicle is currently within a black spot region, the data transfer is postponed since the prediction model cannot be trusted.
%
% Fig. Trade-off: Clustering
%
\begin{figure}[]  	
	\vspace{0cm}
	\centering		  
	\includegraphics[width=1.0\columnwidth]{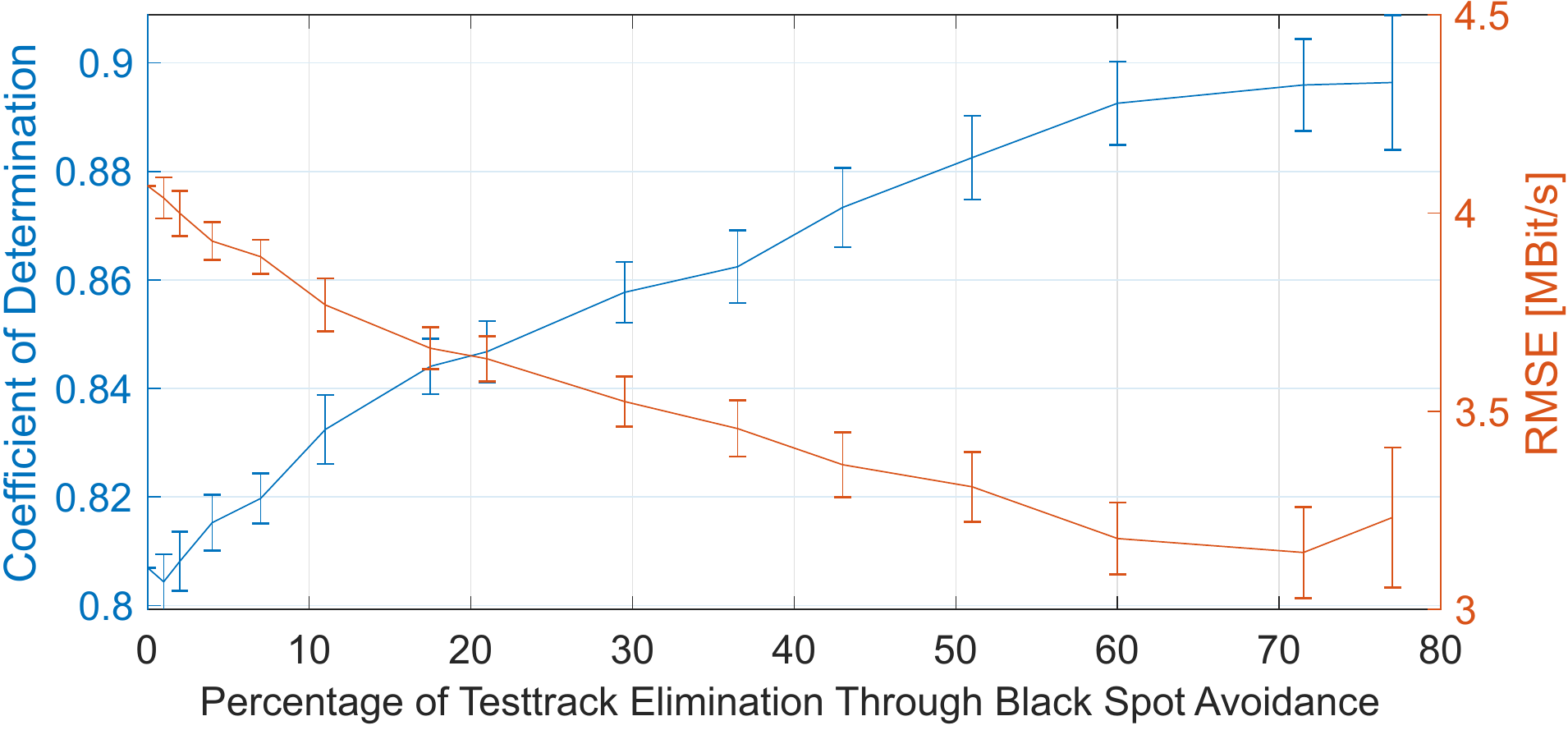}
	\caption{Trade-off between performance improvement of the data rate prediction and tolerable reduction of the  transmission opportunities (\mno{A} uplink).}
	\label{fig:tradeoff_clustering_ul}
	%\vspace{-0.5cm}	
\end{figure}
As a result of this approach, there exists a trade-off between the achievable improvement of the data rate prediction accuracy and a reduction of the usable percentage of the track for performing data transmissions.
Fig.~\ref{fig:tradeoff_clustering_ul} shows the resulting $R^2$ and \ac{RMSE} values with respect to the tolerable percentage of track elimination --- the total spread of the black spot regions over the overall track length --- for the \mno{A} uplink data set of \cite{Sliwa/Wietfeld/2019b}. It can be seen that the reduction of transmission opportunities allows to significantly improve the performance of the prediction model. Where the curves convergence, the model only considers highly reliable connectivity hotspots appropriate for the data transfer. In the following, we allow a maximum track reduction of $20$~ \%.

\section{Methodology} \label{sec:methods}

In this section, an overview about the research methods, tools, and performance metrics is provided. A summary about relevant parameters of the novel transmission scheme is given in Tab.~\ref{tab:parameters}
%
% Tab. Parameters
%
\begin{table}[ht]
	\centering
	%\vspace{-0.1cm}
	\caption{Default parameters of the evaluation setup}
	%\vspace{-0.2cm}
	\begin{tabular}{ll}
		\toprule
		\textbf{Parameter} & \textbf{Value} \\
		
		\midrule
		
		Maximum buffering time $\Delta t_{\max}$ & 120~s \\
		Trade-off factor $w$ & 0.9 \\
		Deadline violation punishment $\Omega$ & -1 \\
		Exploration parameter $\delta$ & 0.1 \\
		Number of clusters $N_{c}$ & 100 \\
		MNO-specific black spot threshold $\text{RMSE}_{\max}$ & 3, 2.25, 2.5 \\
		Periodic data transfer interval $\Delta t$ & 10~s \\
		
		\bottomrule
		
	\end{tabular}
	%\vspace{-0.3cm}
	\label{tab:parameters}
\end{table}

\subsection{Real World Data Acquisition}

%
% Fig. Network model
%
\begin{figure}[]  	
	\vspace{0cm}
	\centering		  
	\includegraphics[width=1.0\columnwidth]{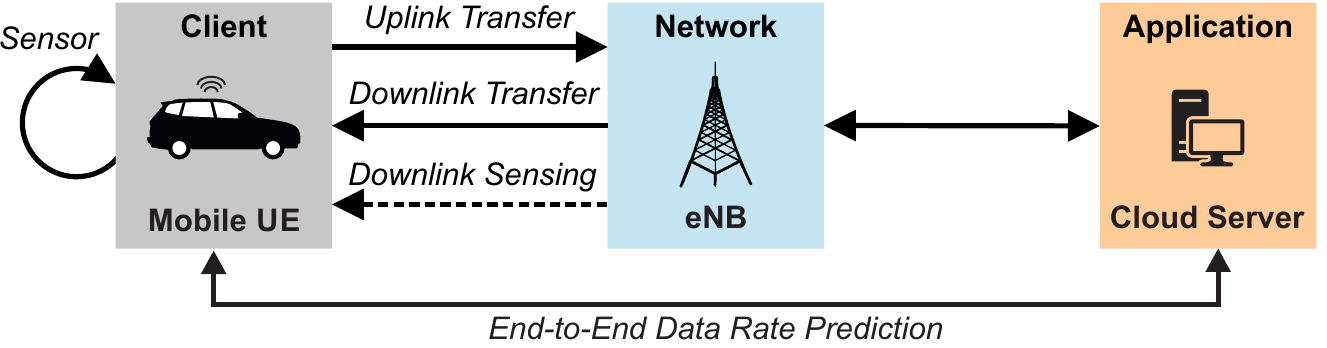}
	\caption{Network model of the real world performance evaluation.}
	\label{fig:networkModel}
	%\vspace{-0.5cm}	
\end{figure}
%
% Scenario
%
For the empirical performance comparison, five test drives are performed in the real world for each of the transmission schemes. Fig.~\ref{fig:networkModel} shows the network model of the evaluation. A virtual sensor application generates 50~kB of sensor data per second. Data transmissions are performed from a moving vehicle through the public \ac{LTE} networks of three German \acp{MNO} in uplink and downlink direction via \ac{TCP}. The evaluations are carried out along a 25~km long evaluation track (Fig.~\ref{fig:map}) which contains highway and suburban regions with varying building densities and speed limitations. In total, 8563 transmissions -- 13.61~GB of transmitted data -- are performed. 
%
% Measurements
%
The passive measurement of the context indicators as well as the active data transmission are performed on an \texttt{Android}-based \ac{UE} (Galaxy S5 Neo, Model SM-G903F) based on a novel application. The latter is provided in an open source way\footnote{Source code available at \url{https://github.com/BenSliwa/MTCApp}}.

\subsection{Performance Indicators}

Within the real world performance comparison in Sec.~\ref{sec:results}, multiple \acp{KPI} are considered which are obtained as follows.

%
% Data rate
%
\textbf{End-to-end data rate}: The evaluation of the achieved data rate is performed at the application level and represents the transmission efficiency of the considered transmission schemes. The actual measurements are performed at a cloud server.

%
% AoI
%
\textbf{\ac{AoI}}: Due to the local buffering process implied by the opportunistic data transfer approach, each transmitted data packet consists of multiple sensor packets. In order to analyze the freshness of the received sensor information, the generation time of the \emph{oldest} sensor packet within the received overall data is considered.

%
% Network resources
%
\textbf{Network resources}: For estimating the number of \acp{PRB} of performed transmissions in a postprocessing step, we revert the procedure described in \cite{Satoda/etal/2020a}. Hereby, the \ac{CQI} measurements are utilized to determine the \ac{MCS} and \ac{TBS} indices from the 3GPP TS 36.213 lookup tables. Based on this information and the measured data rate, the number of \acp{PRB} is inferred.

%
% Power consumption
%
\textbf{Power consumption}: The resulting power consumption of a mobile \ac{UE} is mainly determined by the applied transmission power $P_{\text{TX}}$ which controls the stage of the power amplifiers. Unfortunately, \texttt{Android}-based \acp{UE} do not expose this information to the user space. However, the analysis in \cite{Falkenberg/etal/2018a} has shown that $P_{\text{TX}}$ can be inferred from radio signal measurements since it is highly correlated to distance-dependent indicators such as \ac{RSRP}.
Therefore, we apply the proposed machine learning-based prediction toolchain of \cite{Falkenberg/etal/2018a} to estimate $P_{\text{TX}}$ and determine the transmission-related power consumption based on laboratory measurements of the device-specific power consumption behavior. Additional details about the applied procedure are presented in \cite{Sliwa/etal/2019d}.
We remark that the power consumption is not a major limiting factor for vehicular crowdsensing. Yet, the usage of battery-powered robotic vehicles such as \acp{UAV} for data acquisition in future \acp{ITS} is highly being discussed. In addition, the proposed approach might also be applied in intelligent container systems in smart logistics scenarios.

\subsection{Data-driven Network Simulation}

It is obvious that the inherently huge effort in performing real world test drives makes this method inappropriate for carrying out large scale parameter studies. Therefore, we exploit the computational efficiency of data-driven analysis methods and implement a \ac{DDNS} setup according to \cite{Sliwa/Wietfeld/2019d} for the initial parameter tuning phase.
%
%
%

%
% DDNS
%
In contrast to classical network simulation methods which simulate the behavior of actual communicating entities and their corresponding protocol stacks, \ac{DDNS} relies on replaying previously acquired empirical context traces of the targeted deployment scenario. Hereby, the vehicle is virtually moved on its trajectory and the corresponding context information is lookup up from the measurements. For this purpose, we utilize the available open data set of \cite{Sliwa/Wietfeld/2019b}. The simulation of the end-to-end behavior of the transmission schemes is then performed by a combination of machine learning models:
%
% Components of DDNS
%
\begin{itemize}
	%
	% Deterministic model
	%
	\item Based on the available a priori data set, a \textbf{deterministic} data rate prediction model --- equal to the \ac{RF} method described in Sec.~\ref{sec:prediction} --- is learned and utilized by the agent to opportunistically schedule the data transmissions. However, due to its deterministic nature, identical feature sets will always result in the same prediction results. Contrastingly, in the real world, the predictions will most likely differ from the ground truth measurements due to imperfections of the prediction model.
	%
	% Probabilistic model
	%
	\item For representing this aspect within the simulation process, a \textbf{probabilistic derivation model} is utilized. Through applying \ac{GPR} on the results of the \ac{RF} model (for a visual representation of the different models, see Fig.~\ref{fig:gpr}), a statistical description of the derivations between predictions and measurements is derived. Furthermore, the Bayesian nature of this model class allows to draw sample values from the learned confidence interval. Within the \ac{DDNS} simulation, each deterministic prediction $\tilde{S}(t)$ is converted to a sampled \emph{virtual ground truth} value $\hat{S}(\tilde{S}(t))$ which represents the actual resulting data rate of the corresponding data transmission. Further details about this method are presented in \cite{Sliwa/Wietfeld/2019d}.
\end{itemize}

\subsection{Data Analysis}

For training the prediction models, we utilize the \ac{LIMITS} framework \cite{Sliwa/etal/2020c} which allows to automate low-level machine analysis in \ac{WEKA} \cite{Hall/etal/2009a} and provides automated export of \texttt{C/C++} code of the trained models.
In order to generate the \ac{GPR} models for the \ac{DDNS} setup and for performing the k-means black spot clustering, the \emph{Statistics and Machine Learning Toolbox} of \texttt{MATLAB} is applied.

For analyzing the performance of the machine learning methods, multiple statistical metrics are applied.
The \emph{coefficient of determination} $R^{2}$ is a statistical metric for the \emph{goodness of fit} of the resulting regression model. It is calculated as 
%
% Eq. R^2
%
\begin{equation}
	R^{2} = 1- \frac{\sum_{i=1}^{N}\left(\tilde{y}_{i} - y_{i} \right)^{2}}{\sum_{i=1}^{N}\left(\bar{y} - y_{i} \right)^{2}}\label{eq:r2}
\end{equation}
with $N$ as the number of measurements, $\tilde{y}_{i}$ being the current prediction, $y_{i}$ being the current measurement, and $\bar{y}$ being the mean value of the measurements.

In addition, we consider \ac{MAE} and \ac{RMSE} which are calculated as
%
% Eq. MAE
%
\begin{equation*}
	\text{MAE} = \frac{\sum_{i=1}^{N} | \tilde{y}_{i} - y_{i}|}{N},
	\quad
	\text{RMSE} = \sqrt{\frac{\sum_{i=1}^{N} \left(  \tilde{y}_{i} - y_{i} \right)^2}{N}} .
\end{equation*}
\section{Results} \label{sec:results}
%
% Fig. Trade-off CB
%
\begin{figure}[]  	
	\vspace{0cm}
	\centering		  
	\includegraphics[width=1.0\columnwidth]{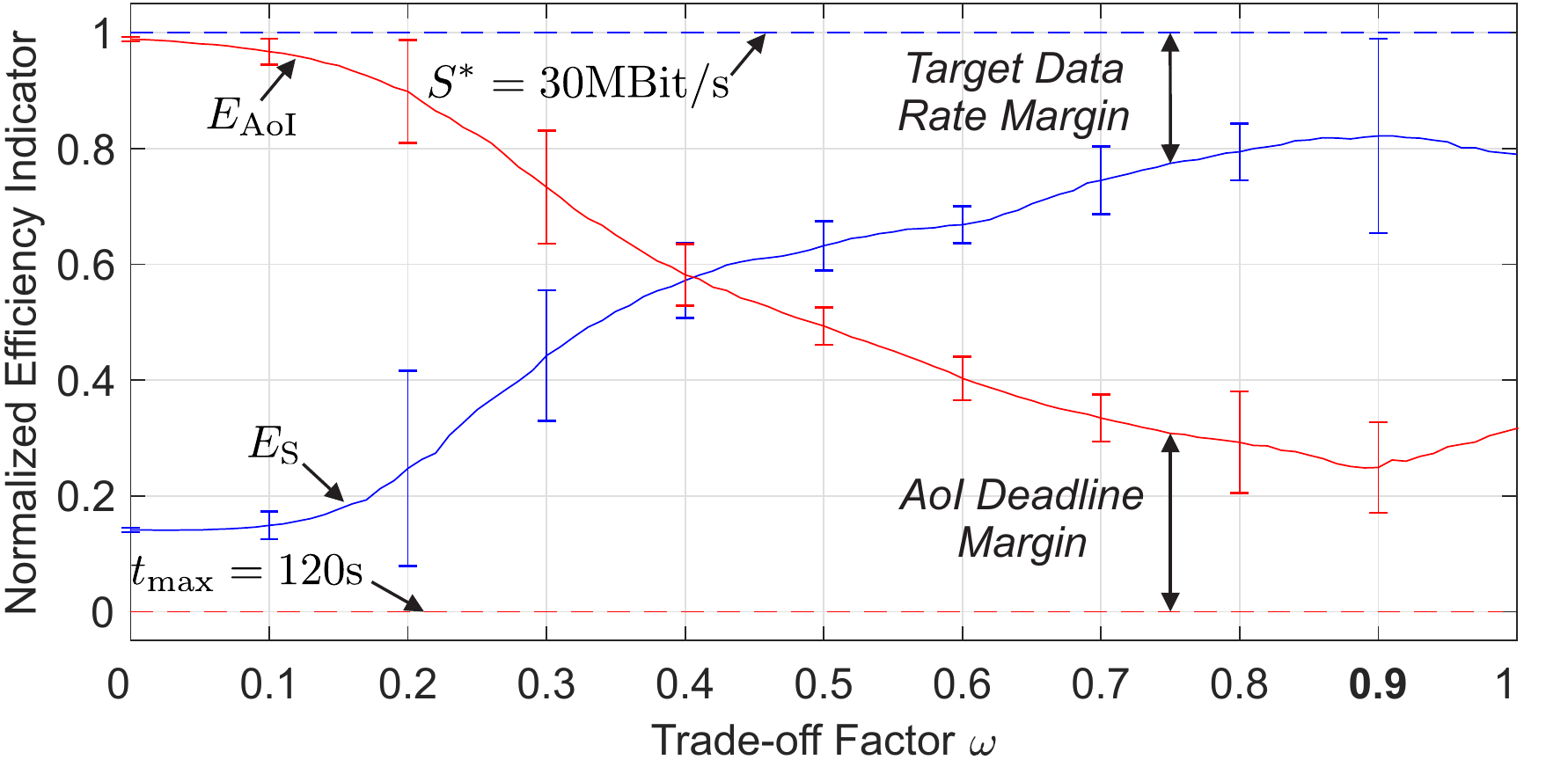}
	\caption{Trade-off between data rate and \ac{AoI} optimization for \mno{A} in uplink direction.}
	\label{fig:tradeoff_cb}
	%\vspace{0cm}	
\end{figure}
%
%
%
%
% Fig. Epochs
%
\begin{figure}[]  	
	\vspace{0cm}
	\centering		  
	\includegraphics[width=1.0\columnwidth]{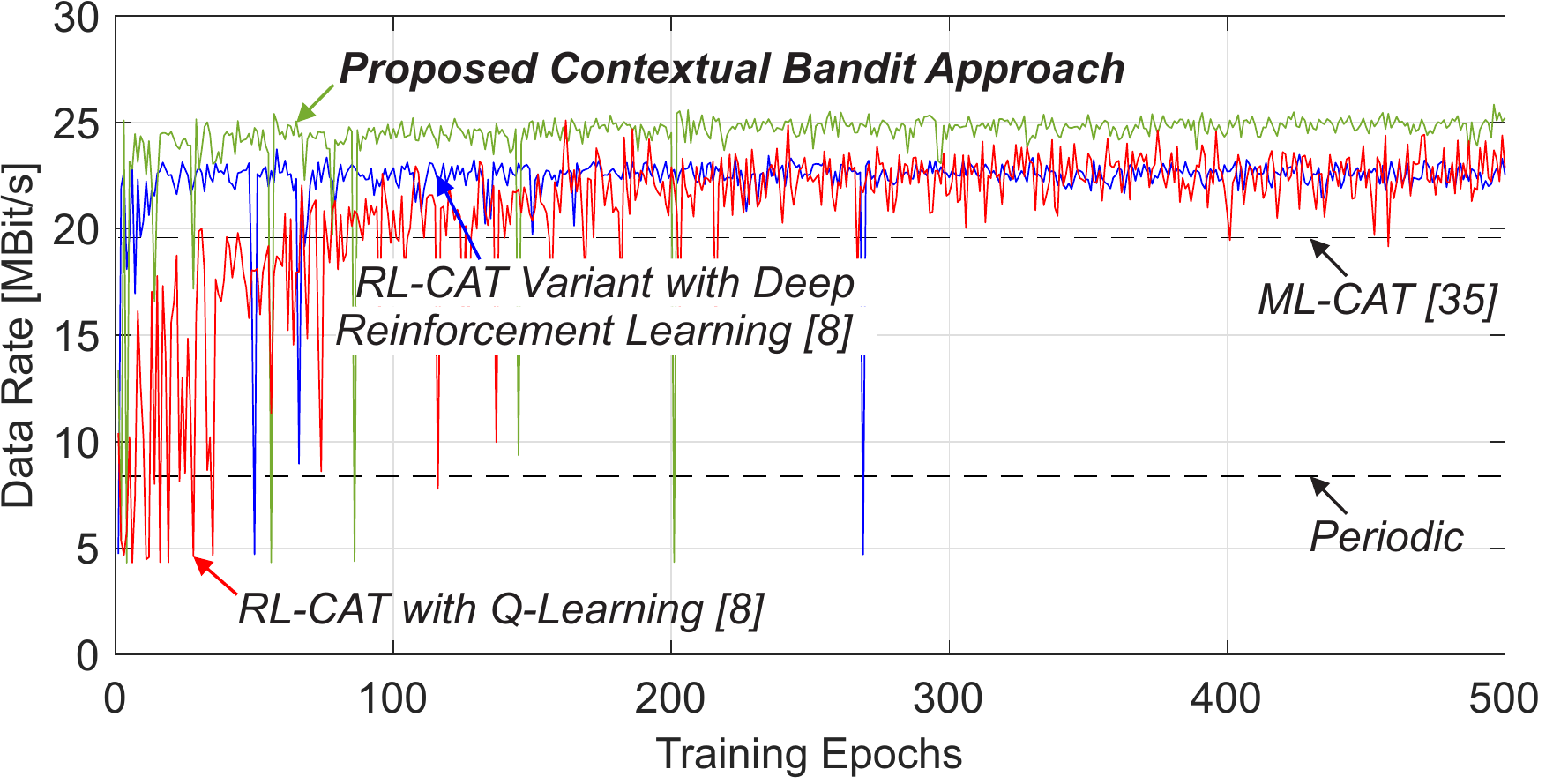}
	\caption{Convergence behavior of the reinforcement learning-enabled transmission schemes. Each epoch corresponds to a virtual test drive evaluation in the \ac{DDNS}.}
	\label{fig:epochs}
	%\vspace{-0.5cm}	
\end{figure}

\renewcommand{\boxWidth}{0.47\textwidth}

In this section, the results for the \ac{DDNS}-based optimization phase as well as for the real world performance analysis are presented and discussed. Within the latter, the novel \proposal method is compared to the existing transmission schemes discussed in Sec.~\ref{sec:continuity}.

%
% Fig. Data rate
%
\renewcommand{\boxWidth}{0.49\textwidth}
\begin{figure*}[] 
	\centering
	\subfloat[Uplink]{\includegraphics[width=\boxWidth]{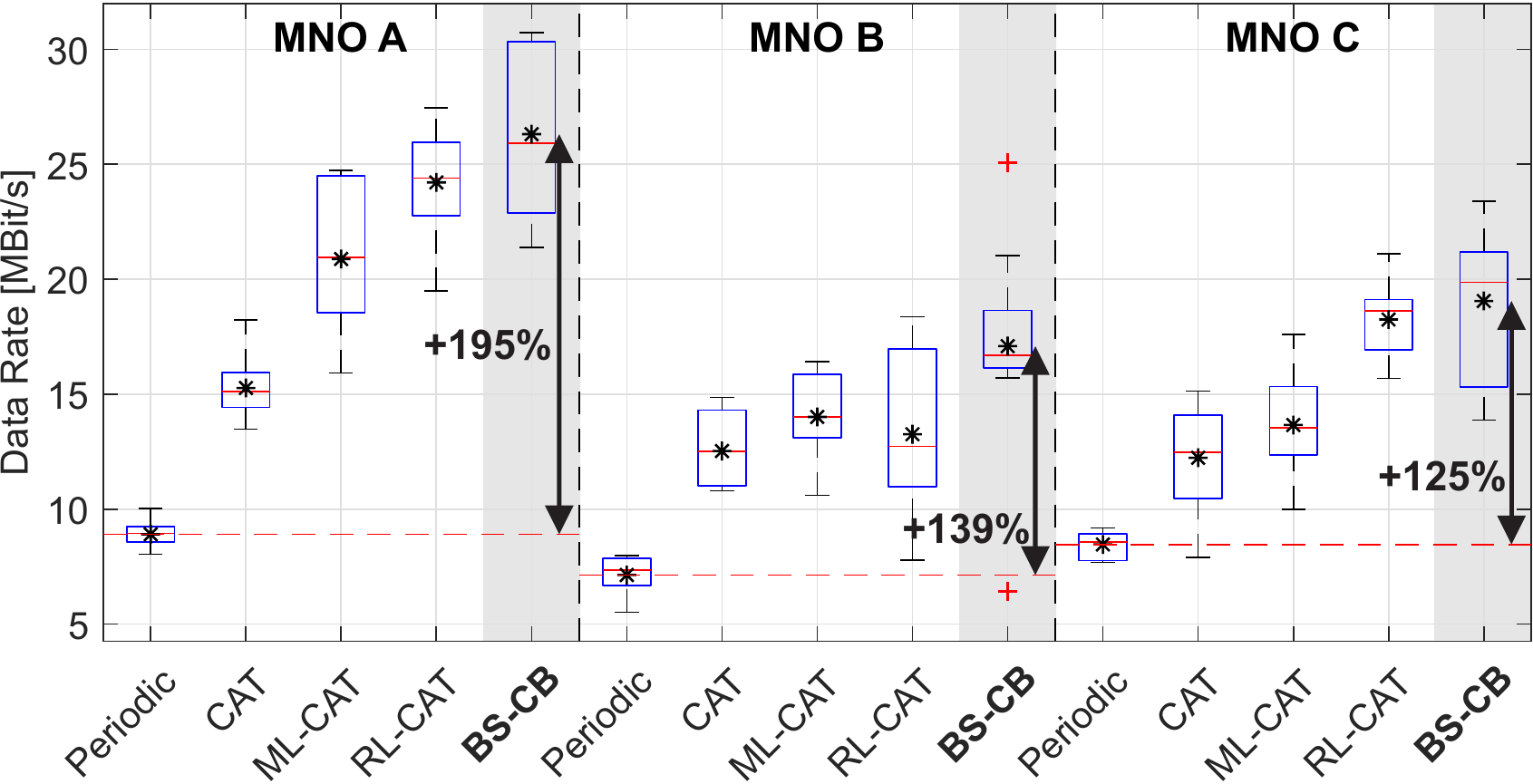}}\hfill
	\subfloat[Downlink]{\includegraphics[width=\boxWidth]{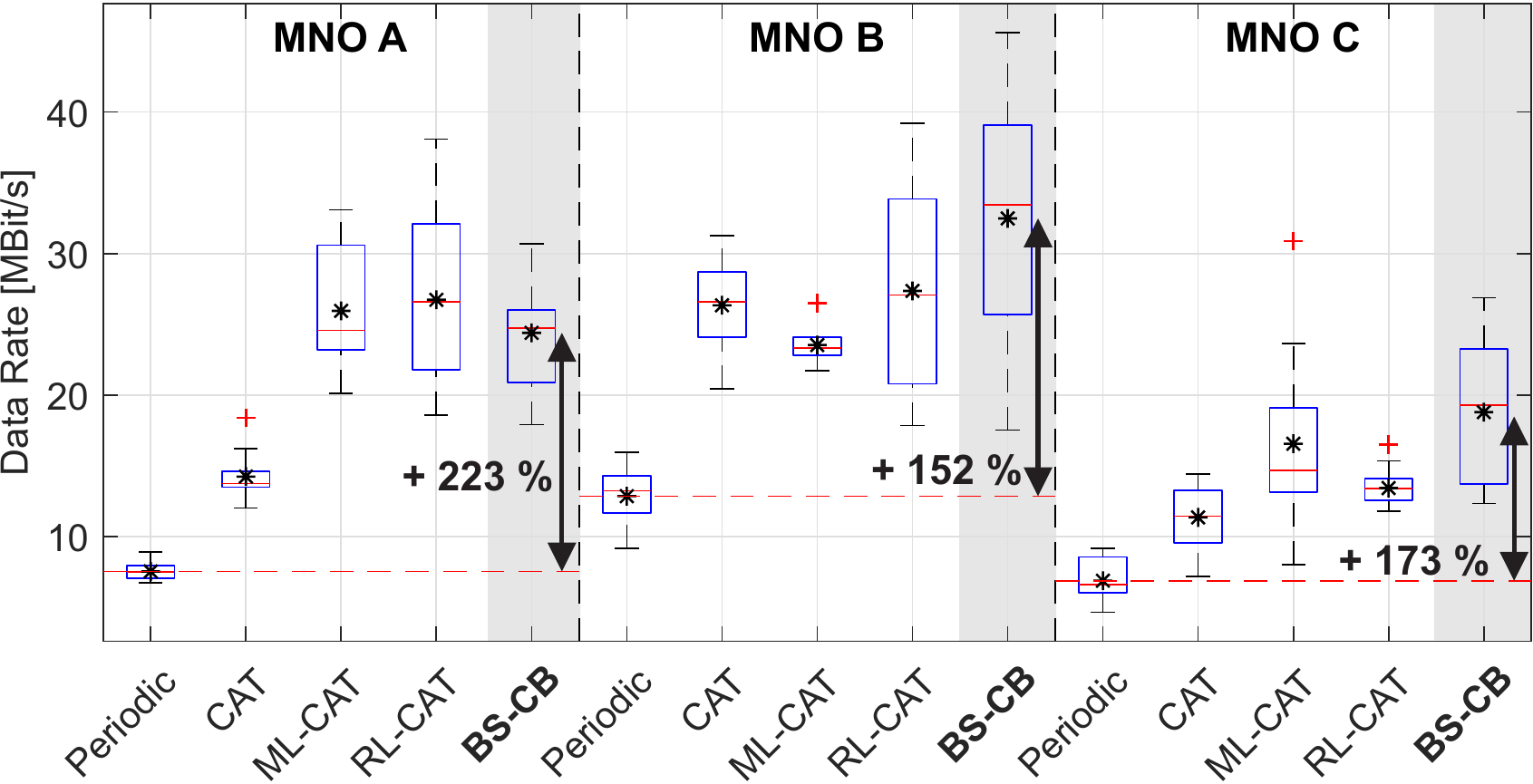}}\hfill		
	\caption{Comparison of the resulting real world data rate in uplink and downlink direction for the considered transmission schemes and \acp{MNO}.}
	\label{fig:cat_data_rate}
	%\vspace{-0.7cm}
\end{figure*}
%
%
%
%
% Fig. Load
%
\renewcommand{\boxWidth}{0.49\textwidth}
\begin{figure*}[] 
	\centering
	\subfloat[Uplink]{\includegraphics[width=\boxWidth]{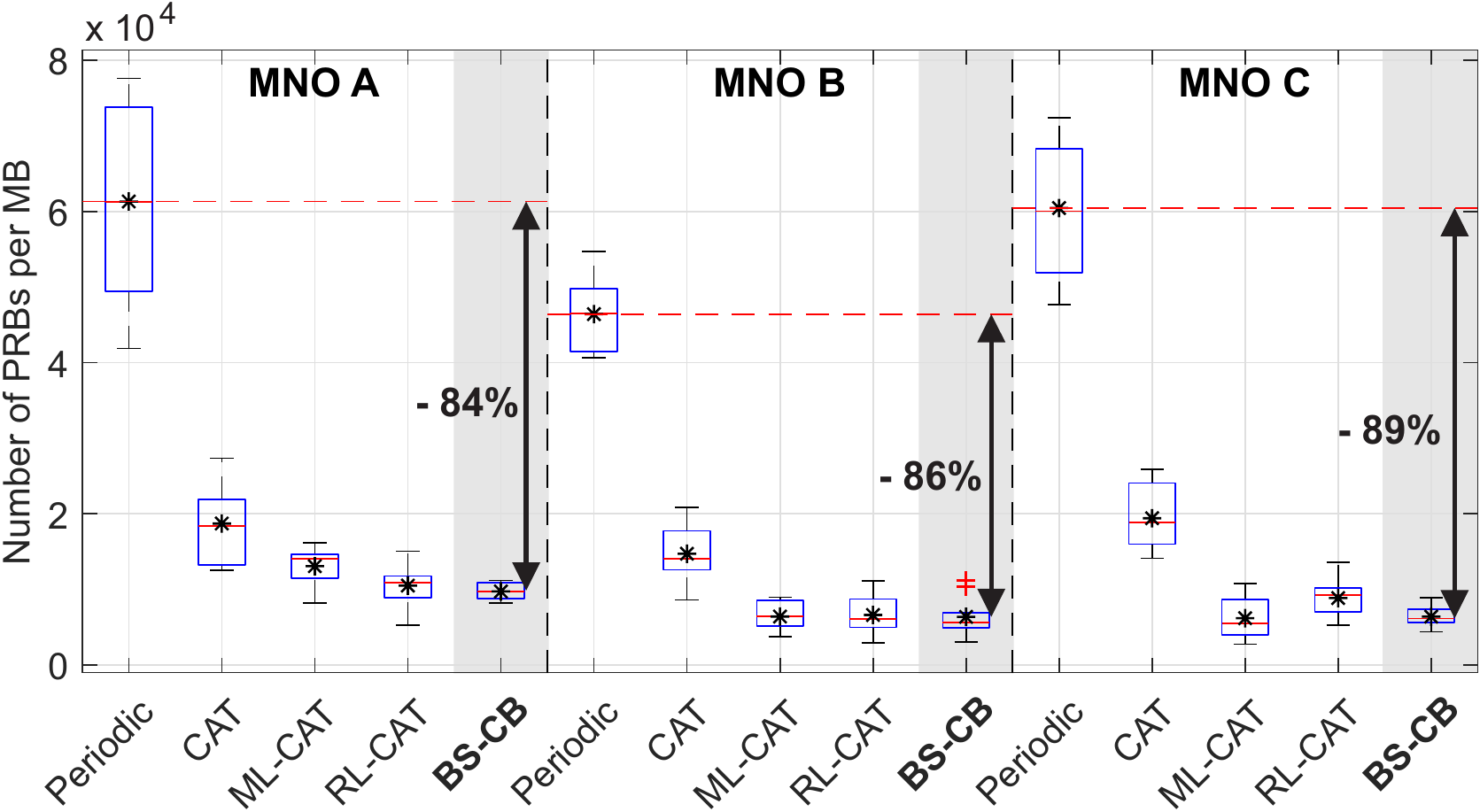}}\hfill
	\subfloat[Downlink]{\includegraphics[width=\boxWidth]{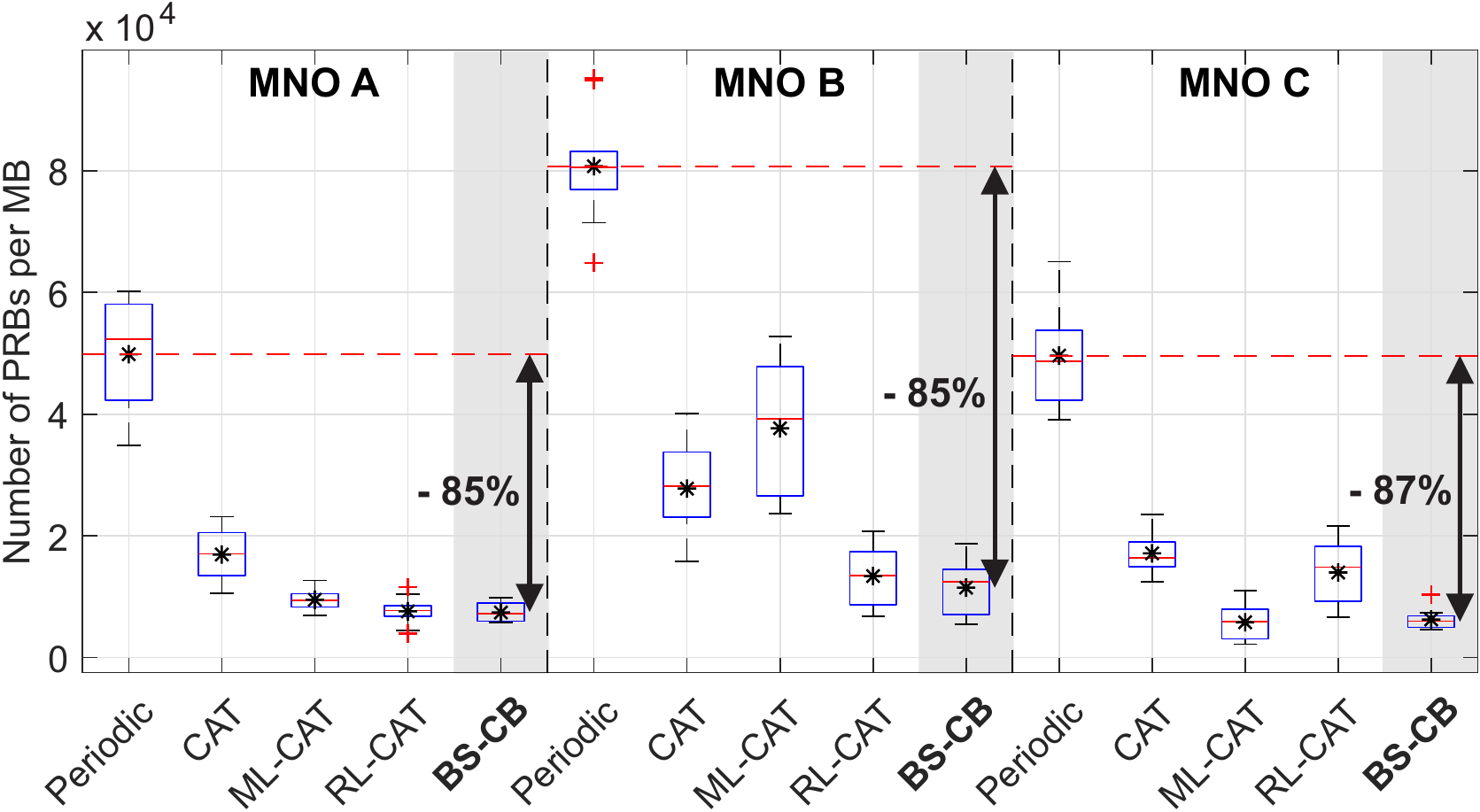}}\hfill		
	\caption{Comparison of the resulting real world resource efficiency in uplink and downlink direction for the considered transmission schemes and \acp{MNO}.}
	\label{fig:cat_load}
	%\vspace{-0.7cm}
\end{figure*}
%
% DDNS
%
\subsection{DDNS-based Parameter Optimization}

%
% Trade-off
%
As discussed in Sec.~\ref{sec:rl}, opportunistic data transfer is subject to a fundamental trade-off between data rate and \ac{AoI} optimization: In order to improve the end-to-end data rate, the transmission schemes will rather prefer larger packets which are then transmitted within connectivity hotspots. As a result of the local buffering, the \ac{AoI} is increased.
For the further analysis of this effect, two efficiency indicators are defined:
%
% Efficiency indicators
%
\begin{itemize}
	%
	% Data rate efficiency
	%
	\item The \textbf{data rate efficiency} $E_{s} = \bar{S} / S^{*}$ is used to analyze how good the average data rate $\bar{S}$ approaches the target data rate $S^{*}$.
	%
	% AoI efficiency
	%
	\item The \textbf{\ac{AoI} efficiency} $E_{\text{AoI}} = 1 - \bar{\Delta} t / \Delta t_{\max}$ represents a measure for the margin between the average \ac{AoI} and the application-specific deadline  $\Delta t_{\max}$ of the age of the sensor data.
\end{itemize}

The fundamental trade-off between data rate optimization and \ac{AoI} optimization which is controlled via the trade-off factor $w$ is shown in Fig.~\ref{fig:tradeoff_cb}.
%
% General
%
It can be seen that the resulting data rate can be improved by transmitting larger data packets based on a larger value of $w$. However, this is achieved through a higher buffering time of the acquired sensor data packets which increases the \ac{AoI} of the data packets.
%
% Conclusion: Value selection for w
%
In the following, we focus on data rate optimization and apply $w=0.9$ within all considered evaluations.

%
% Epochs
%
Although the reinforcement learning mechanisms can theoretically be learned online in the field, we apply an offline training approach based on \ac{DDNS} in order to ensure that the real world evaluations are performed with a converged system.
Hereby, we replay previously acquired empirical context traces --- which are referred to as \emph{epochs} --- and apply the novel reinforcement learning-based transmission schemes. The resulting data rate behavior is shown in Fig.~\ref{fig:epochs}. As references, we consider the Q-learning-based \ac{RL-CAT} and a deep reinforcement learning variant of the latter which applies an \ac{ANN} configuration according to Sec.~\ref{sec:prediction} for the data rate prediction.
%
% CB
%
It can be seen that the contextual bandit-based method achieves the highest absolute data rate and reaches a converged system state early after 200 epochs. The remaining error floor is caused by the imperfections of the data rate prediction model.
%
% Q-Learning
%
For \ac{RL-CAT}, both variants achieve a similar performance level --- about $2.5$~MBit/s less than \proposal --- of the converged methods. However, it can be seen that the deep reinforcement learning variant achieves a faster convergence behavior than the simple Q-learning approach.

%
% Fig. Power consumption
%
\begin{figure}[]  	
	\vspace{0cm}
	\centering		  
	\includegraphics[width=1.0\columnwidth]{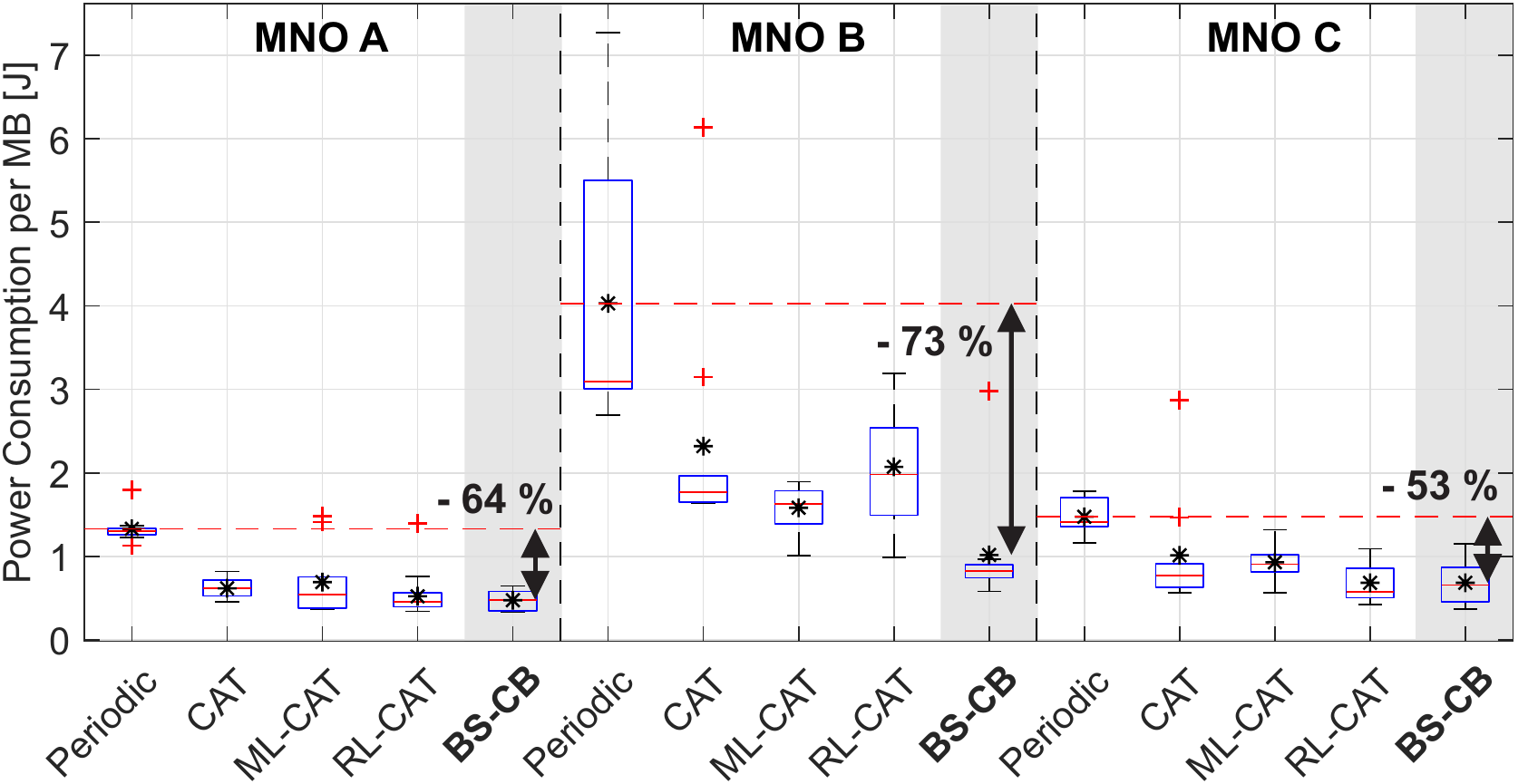}
	\caption{Transmission-related real world uplink power consumption of the mobile \ac{UE}.}
	\label{fig:cat_power_consumption}
	%\vspace{-0.7cm}	
\end{figure}
%
%
%
%
% Fig. AoI
%
\begin{figure*}[] 
	\centering
	\subfloat[Uplink]{\includegraphics[width=\boxWidth]{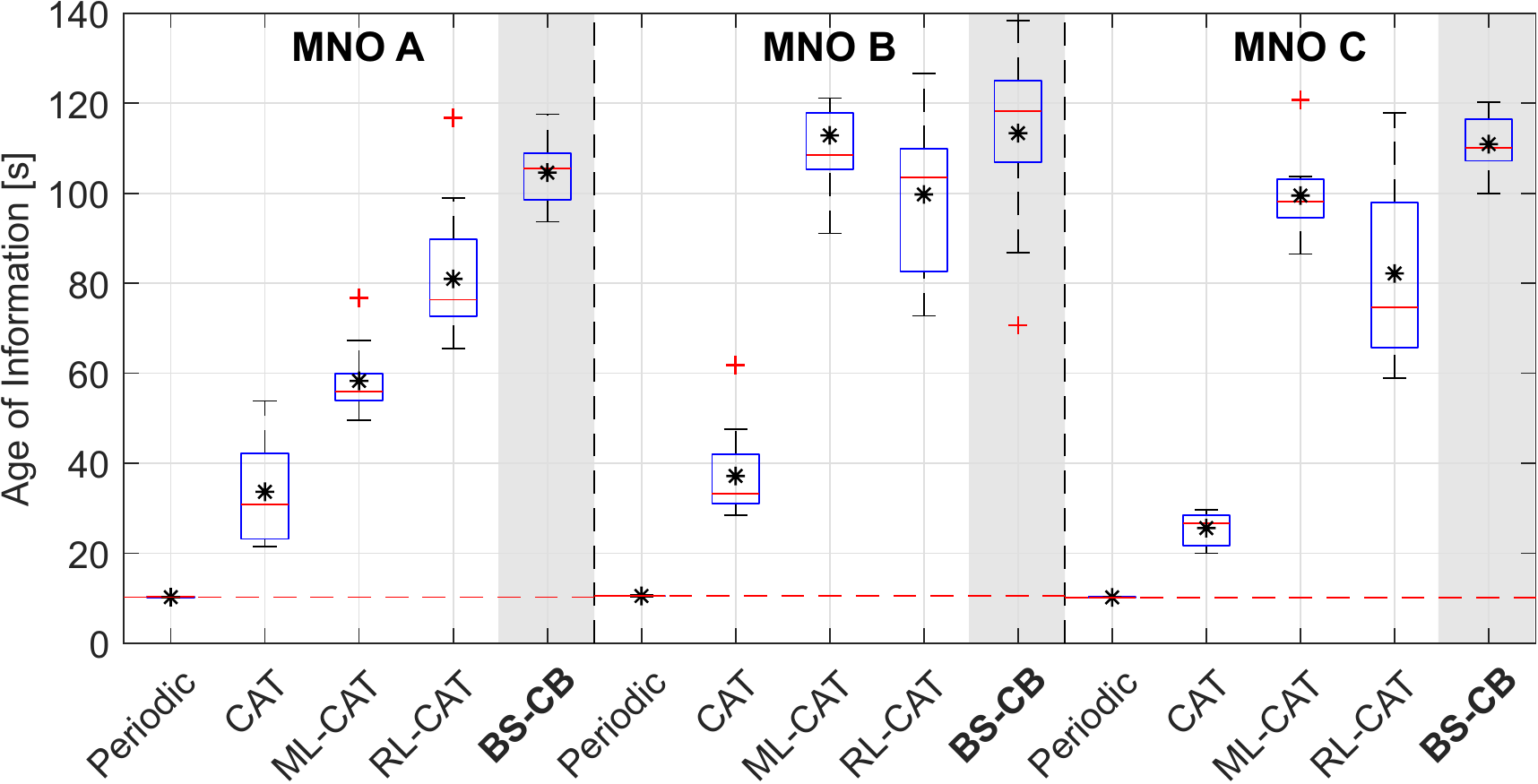}}\hfill
	\subfloat[Downlink]{\includegraphics[width=\boxWidth]{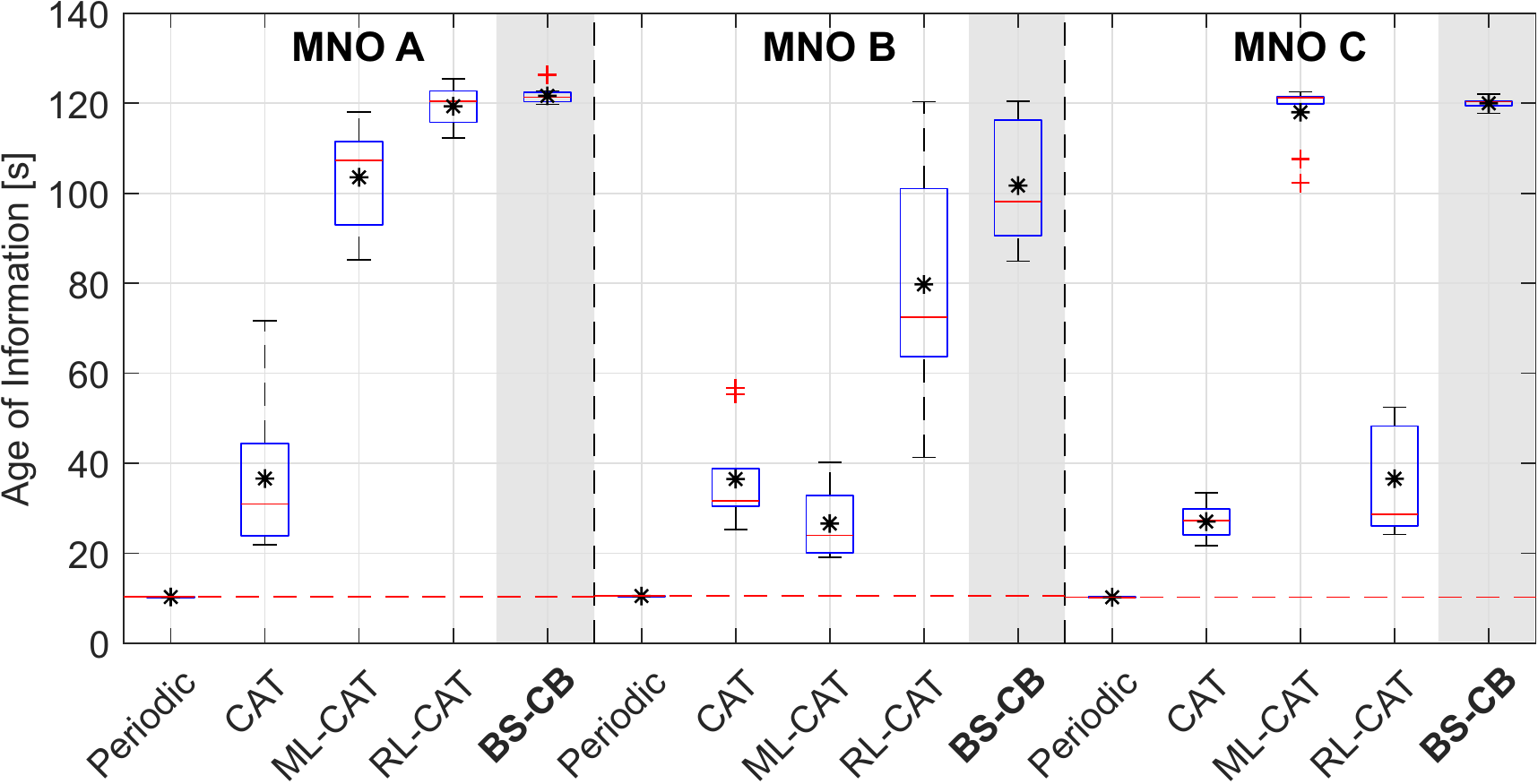}}\hfill		
	\caption{Comparison of the resulting real world \ac{AoI} of the sensor data packets.}
	\label{fig:cat_aoi}
	%\vspace{-0.5cm}
\end{figure*}
%
%
%

%
% REAL WORLD PERFORMANCE COMPARISON
%
\subsection{Real Wold Performance Comparison}

The configured and converged transmission schemes are now applied in a real world evaluation and compared to existing transmission approaches.

The resulting data rate of the different transmission schemes is shown in Fig.~\ref{fig:cat_data_rate} for uplink and downlink direction.
%
% Periodic -> CAT -> ML-CAT
%
A clear trend of continuous improvement over the different evolution stages can be observed: Although already the \ac{SINR}-based \ac{CAT} method is able to achieve significant improvements in comparison to the periodic data transfer approach, the introduction of the machine learning-based data rate prediction metric by \ac{ML-CAT} leads to a significant boost which is the result of a more reliable way of accessing the channel behavior.
Finally, it can be seen that the reinforcement learning-based decision making outperforms the previously considered heuristic approaches. Hereby, data rate improvements up to $195$~\% in uplink and up to $223$~\% in downlink direction are achieved by the proposed \proposal method.
%
% Uplink vs Downlink
%
In the downlink, the differences between the opportunistic transmission approaches are less distinct since the downlink performance is more determined by the network congestion than the radio channel conditions \cite{Bui/etal/2017a}.

A comparison of the resulting network ressource efficiency (represented by the amount of \acp{PRB} per transmitted {MB}) is shown in Fig.~\ref{fig:cat_load}. It can be seen that all opportunistic data transfer approaches are able to massively reduce --- by 84~\% to 89~\% --- the amount of occupied network resources for all \acp{MNO} in both transmission directions.
One of the main reasons for this behavior is the explicit exploitation of \emph{connectivity hotspot} situations. Here, the robust channel conditions allow to apply higher \acp{MCS} for the actual data transfer. Again, it can be seen that the more advanced evolution stages of the \ac{CAT} approach allow to identify these favorable transmission opportunities in a more reliable way.
As a conclusion, the apparently selfish goal of data rate optimization contributes to improving the intra-cell coexistence: Since the limited \acp{PRB} are only occupied for small amounts of time, they are freed early and are available for being allocated by other cell users.

The resulting uplink power consumption of the mobile \ac{UE} is shown in Fig.~\ref{fig:cat_power_consumption}. Since the opportunistic data transmission schemes aim to exploit connectivity hotspots, they implictly increase the average \ac{RSRP} at the transmission time which is highly correlated to the applied transmission power.
As discussed in \cite{Falkenberg/etal/2018a}, the latter is the major impact factor for the uplink power consumption since it controls the state of the different power amplifiers of the \ac{UE}. Therefore, the \ac{RSRP} optimization leads to a massive improvement of the observed power consumption. Here, \proposal is able to reduce the latter between $-53$~\% and $-73$~\%.
% 
% MNO B
%
For \mno{B}, it can be observed that the general level of the uplink power consumption is much higher than for the other \acp{MNO}. However, this phenomenon is caused by network planning-related aspects of the operator: In the considered evaluation scenario, the average distance to the \acp{eNB} is much higher than for the other \acp{MNO}. As a consequence of the resulting \ac{RSRP} reduction --- the average \ac{RSRP} for \mno{B} is $-97.64$~dBm, $-89.61$~dBm for \mno{A}, and $-88.03$~dBm for \mno{C} --- the mobile \ac{UE} applies a higher transmission power to compensate the path loss effects.

Although the considered opportunistic data transfer approaches are able to achieve massive improvements in data rate, network resource efficiency, and uplink power consumption, the price to pay is a significant increase in the \ac{AoI} of the sensor data packets. Fig.~\ref{fig:cat_aoi} shows a comparison of the resulting \ac{AoI} values for the different transmission schemes, \acp{MNO}, and transmission directions.
%
% General
%
The plots show that this effect is more distinct for the machine learning approaches which detect favorable transmission opportunities more reliably through considering the radio channel quality, protocol-related aspects and partially also the network load. 
In contrast to that, the highly dynamic behavior of the \ac{SINR} (see Fig.~\ref{fig:sinr_trace}) leads to a higher transmission probability for the regular \ac{CAT} method which results in a comparably low \ac{AoI}.
However, based on the parameter $\Delta t_{\max}$, the tolerable \ac{AoI} can be configured with respect to the application requirements.
%
% Fig. Indicators vs tMax
%
\begin{figure}[]  	
	\vspace{0cm}
	\centering		  
	\includegraphics[width=1.0\columnwidth]{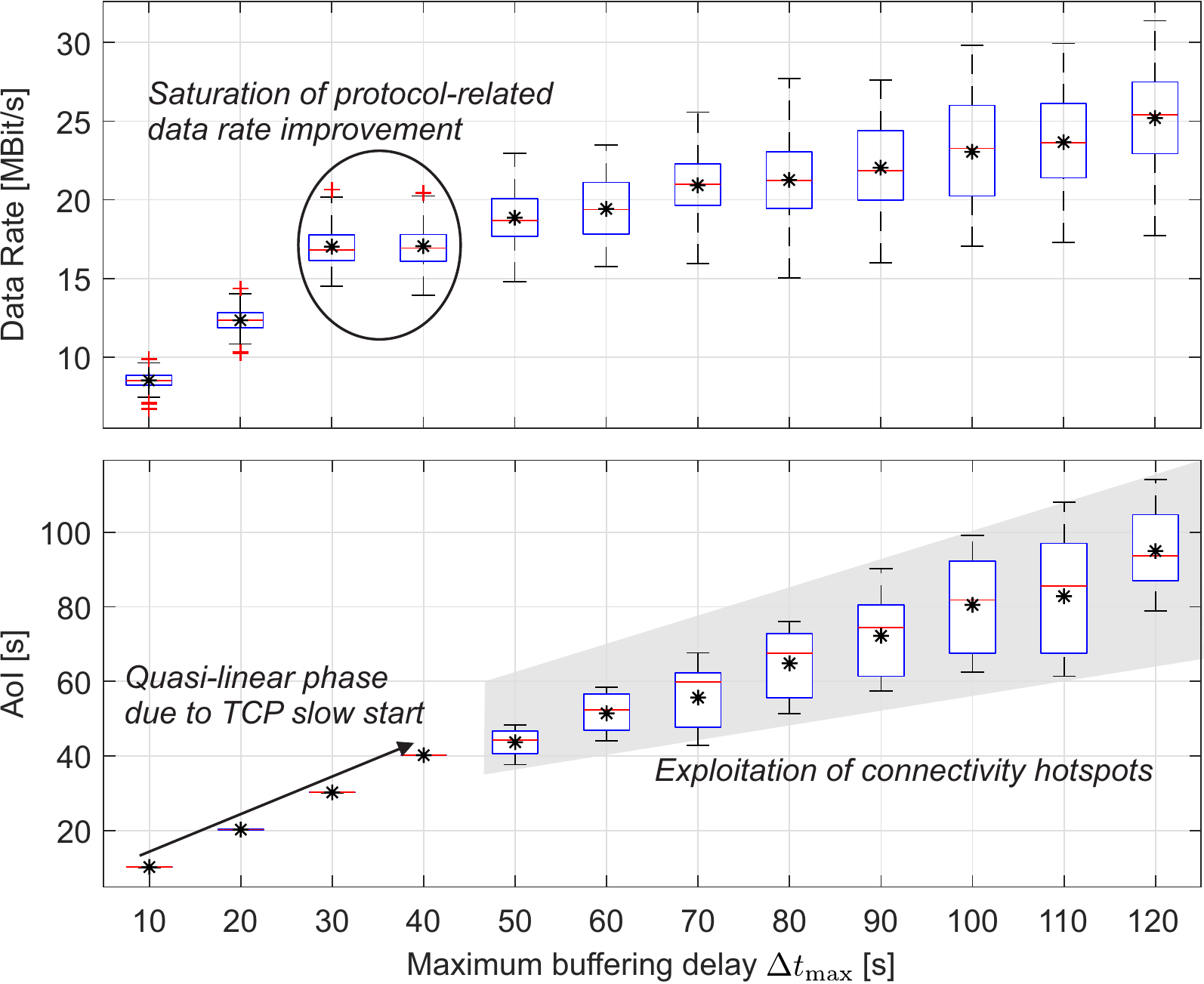}
	\caption{Impact on the application-specific deadline $\Delta t_{\max}$ on the resulting data rate and \ac{AoI}.}
	\label{fig:aoi_vs_tmax}
	%\vspace{-0.7cm}	
\end{figure}
The impact of different values of $\Delta t_{\max}$ on the resulting \proposal data rate and the \ac{AoI} of sensor data is shown in Fig.~\ref{fig:aoi_vs_tmax}.
%
% Quasi-linear phase
%
For small values of $\Delta t_{\max}$, a quasi-linear dependency to the latter can be observed. In this phase, the behavior of the transmission scheme is dominated by protocol effects such as \ac{TCP} slow start. However, a saturation of the data rate improvement is reached at $\Delta t_{\max}=30$~s. 
%
% Opportunistic phase
% 
Afterwards, the actual opportunistic behavior starts which exploits the vehicle's mobility behavior for postponing data transmissions to more robust radio channel conditions where a better resource efficiency can be achieved.

%
% Fig. Spider
%
\begin{figure}[]  	
	\vspace{0cm}
	\centering		  
	\includegraphics[width=1.0\columnwidth]{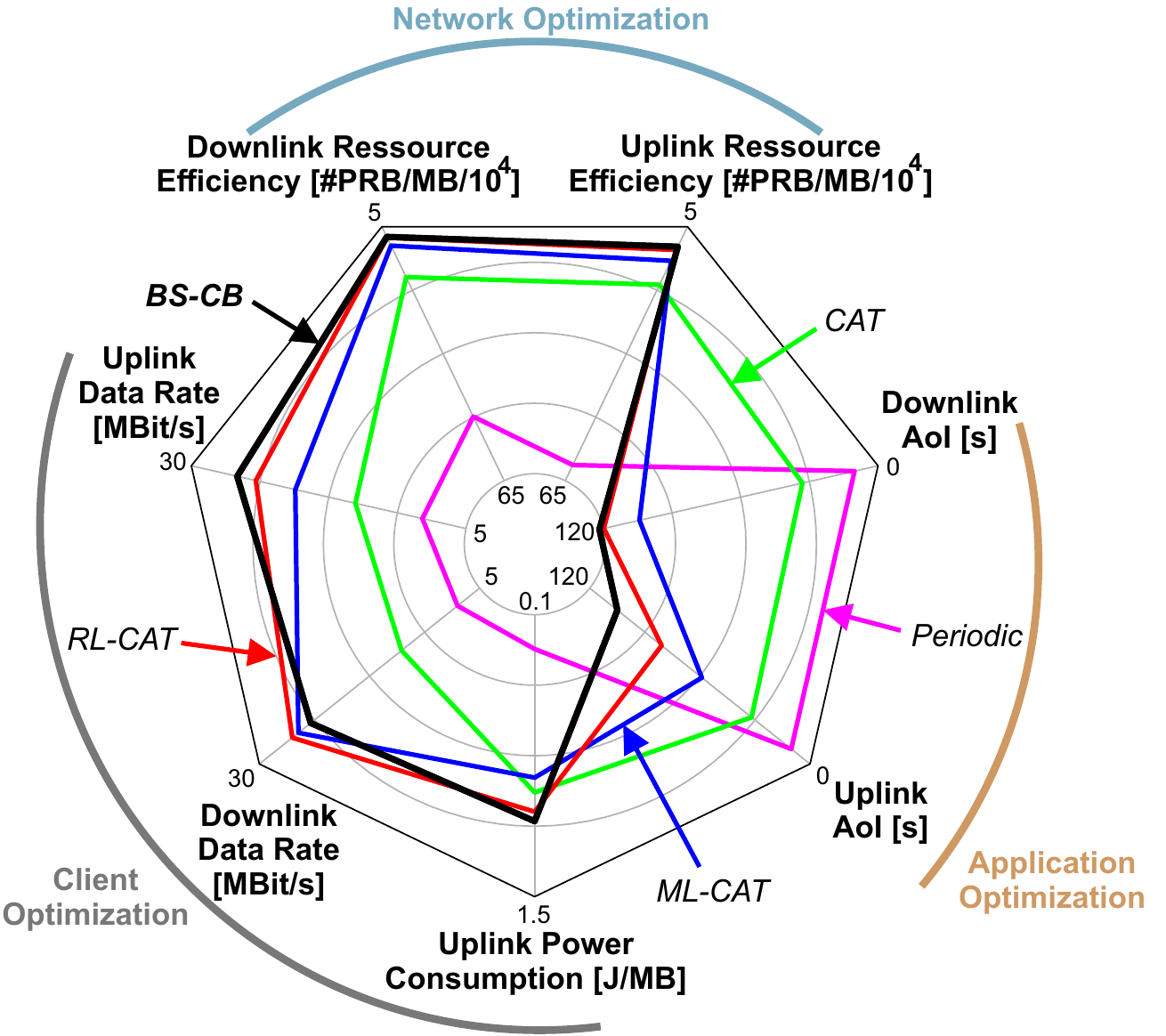}
	\caption{Summary: Comparison of the average behavior of different performance indicators for the opportunistic data transfer methods in the cellular network of \mno{A}. The axis orientations are chosen such that a large footprint represents a better performance.}
	\label{fig:spider}
	%\vspace{-0.7cm}	
\end{figure}
As a summary, Fig.~\ref{fig:spider} shows a spider plot which compares the mean results of all considered performance indicators for the different opportunistic data transfer methods in the network of \mno{A}. The axis orientation have been chosen such that a larger footprint corresponds to a better performance.
It can be seen that all non-periodic approaches focus on optimizing the network and client domain at the expense of the application domain.
Although the proposed \proposal achieves a slightly better overall performance than \ac{RL-CAT}, the major differences can be observed between different categories and less between actual transmission schemes: The highest gains are achieved by the hybrid machine learning approaches that utilize data rate prediction and reinforcement learning-based autonomous decision making.

%
% ONLINE LEARNING
%
\subsection{Online Learning for Self Adaptation to Concept Drift}

%
% Introduction
%
The results of the real world performance evaluation have shown that the client-based machine learning-enabled transmission schemes are able to achieve significant improvements in comparison to existing approaches. 
%
% Concept drift
%
However, changes in the network (e.g., new resource schedulers in the network infrastructure) might lead to a \emph{concept drift} \cite{Gama/etal/2014a} situation where the interplay of the considered features experiences a significant change. Although the application of reinforcement learning allows to further optimize the autonomous decision making during the live evaluations, the data rate prediction model is trained in a static way and might experience a significant reduction of the prediction accuracy. 
%
% Online learning
%
While it is possible to periodically re-train the prediction model, a better approach is the application of \emph{online learning} in order to enable self adaption to the changed environment conditions. 
With respect to the edge intelligence classification scheme of \cite{Zhou/etal/2019a}, the integration of online learning would migrate the transmission scheme to \emph{level 6: all on-device} where training and inferencing are run completely locally.
%
%
%

%
% Setup
%
Since it is not possible for us to cause concept drift in the public cellular network, we virtually create a situation where the network behavior spontaneously changes significantly. For this purpose, we pre-train a prediction model on the uplink data set of one \ac{MNO} and analyze its online adaption to the data set of a different operator.
Although online learning variants of \acp{RF} exist --- e.g., Mondrian Forests \cite{Lakshminarayanan/etal/2014a} --- we apply an \ac{ANN} model for this purpose since this model class inherently supports incremental learning.
%
% Data split
%
For the proof-of-concept experiment, a data split is applied: 80~\% of the \ac{MNO}-specific data set $\mathcal{D}$ is used as the training set $\mathcal{D}_{\text{train}}$ and the remaining data forms the test set $\mathcal{D}_{\text{test}}$. Initially, the \ac{ANN} is pre-trained on the training data of \mno{A}, and then incrementally updated with the training data of \mno{B}. For both operators, the \ac{RMSE} on the corresponding test sets is analyzed.

%
% ANN setup / Data split 
%
The \ac{ANN} is set up according to Sec.~\ref{sec:prediction}. For the incremental learning, a minibatch of $32$ elements is applied. Hereby, the measurements are buffered locally until the buffer size is equal to $32$. Afterwards, the weights of the \ac{ANN} are updated and the buffer is cleared. The resulting \ac{RMSE} on the test sets of both network operators is shown in Fig.~\ref{fig:online_learning_epochs}. Four different characteristic phases can be identified:

%
% Fig. Online learning epochs
%
\begin{figure}[]  	
	\vspace{0cm}
	\centering		  
	\includegraphics[width=1.0\columnwidth]{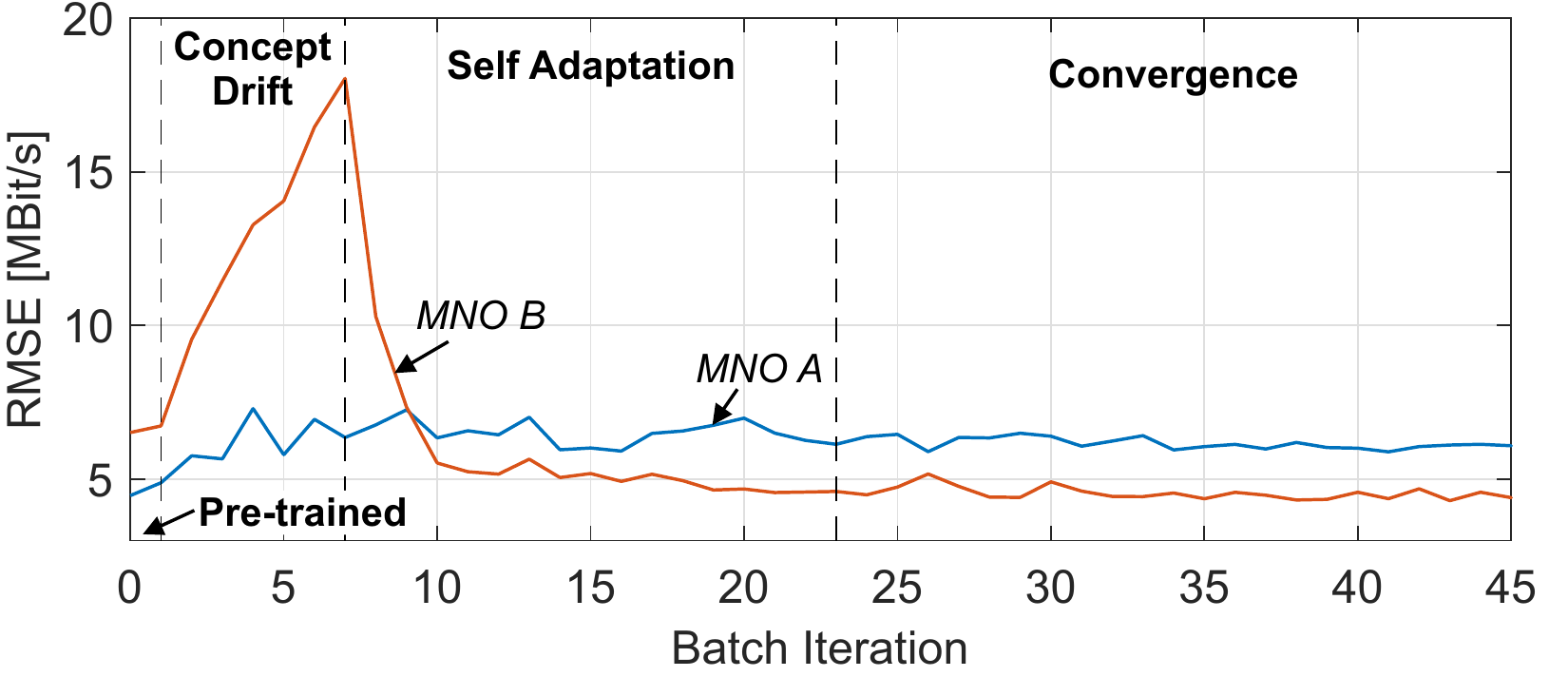}
	\caption{Self adaption of the data rate prediction model to concept drift: An \ac{ANN} model is pre-trained on the uplink data of \mno{A} and then incrementally updated with measurements of \mno{B}.}
	\label{fig:online_learning_epochs}
	%\vspace{-0.7cm}	
\end{figure}
%
%
%
%
% Fig. Black spot statistics
%
\begin{figure*}[] 
	\centering
	\subfloat[]{\includegraphics[width=0.33\textwidth]{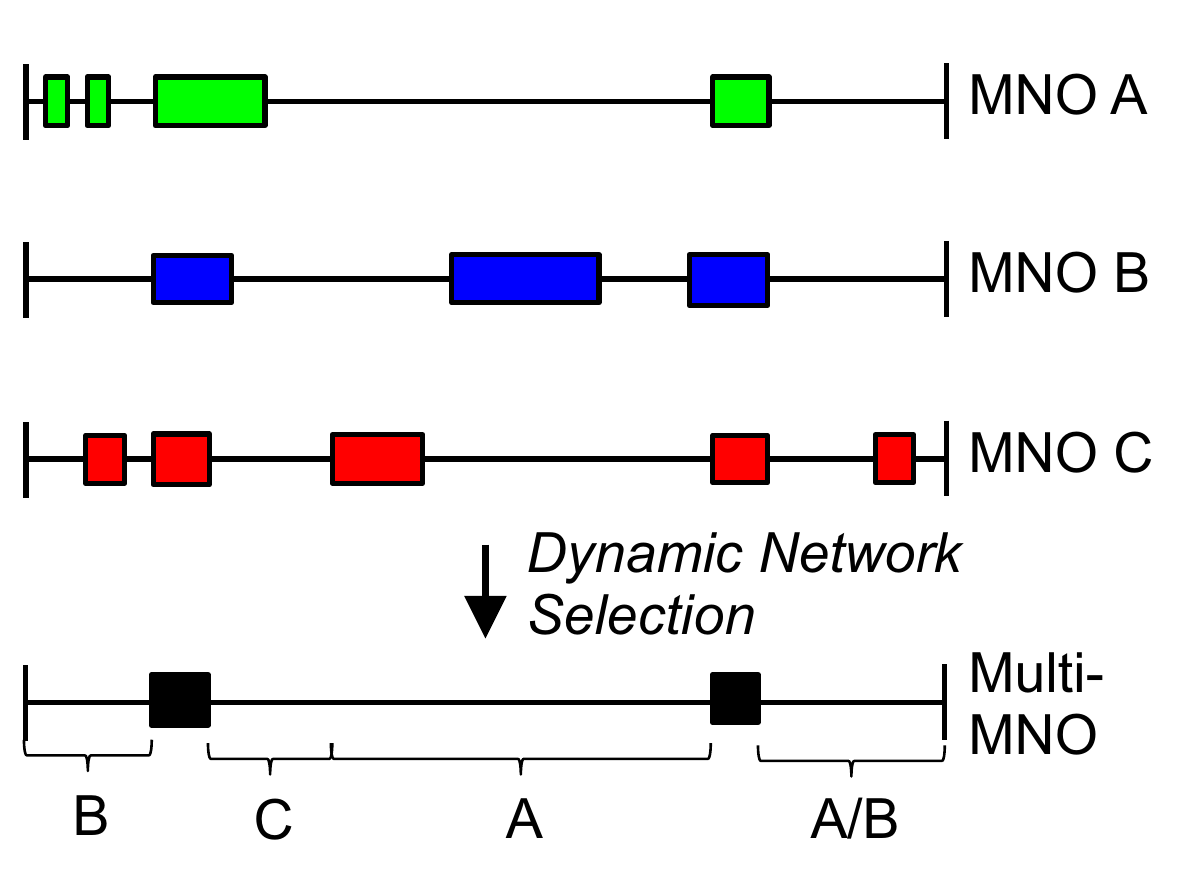}}\hfill
	\subfloat[]{\includegraphics[width=0.33\textwidth]{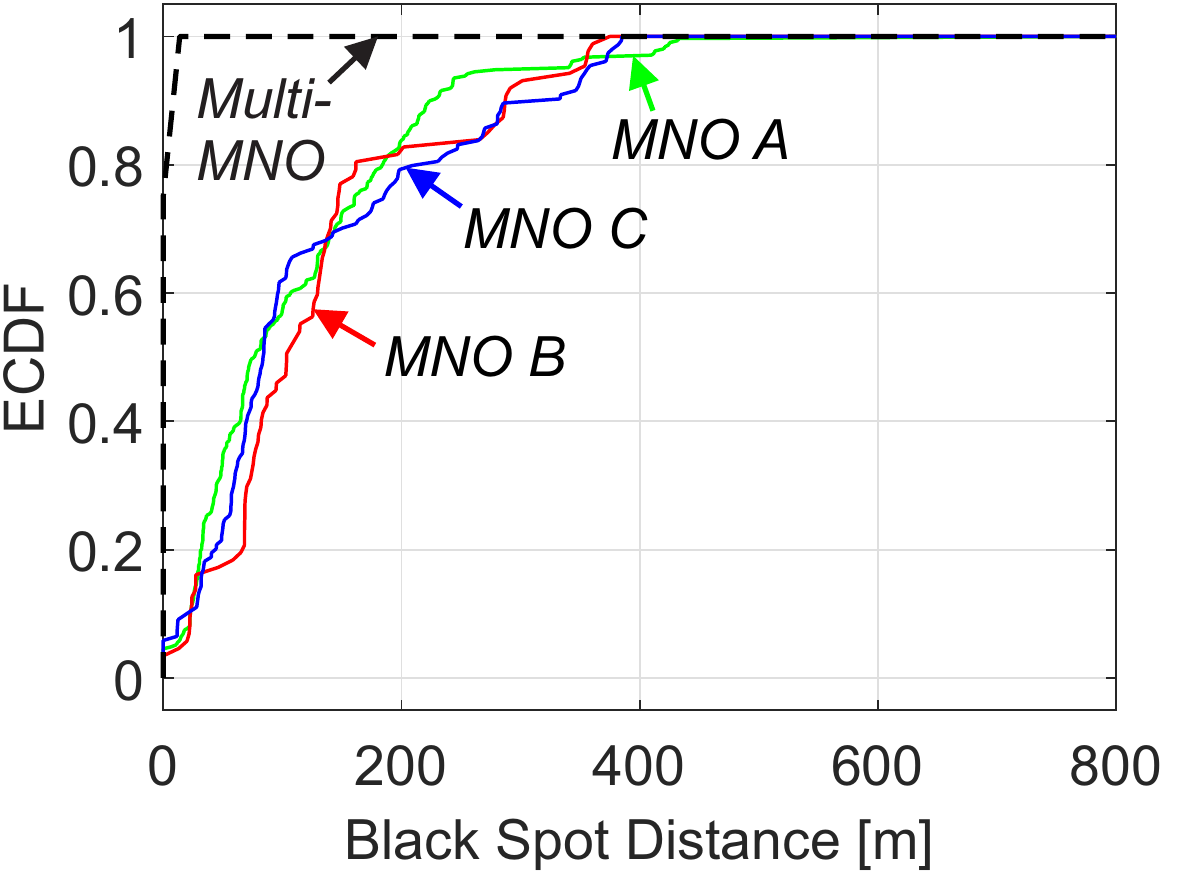}}\hfill
	\subfloat[]{\includegraphics[width=0.33\textwidth]{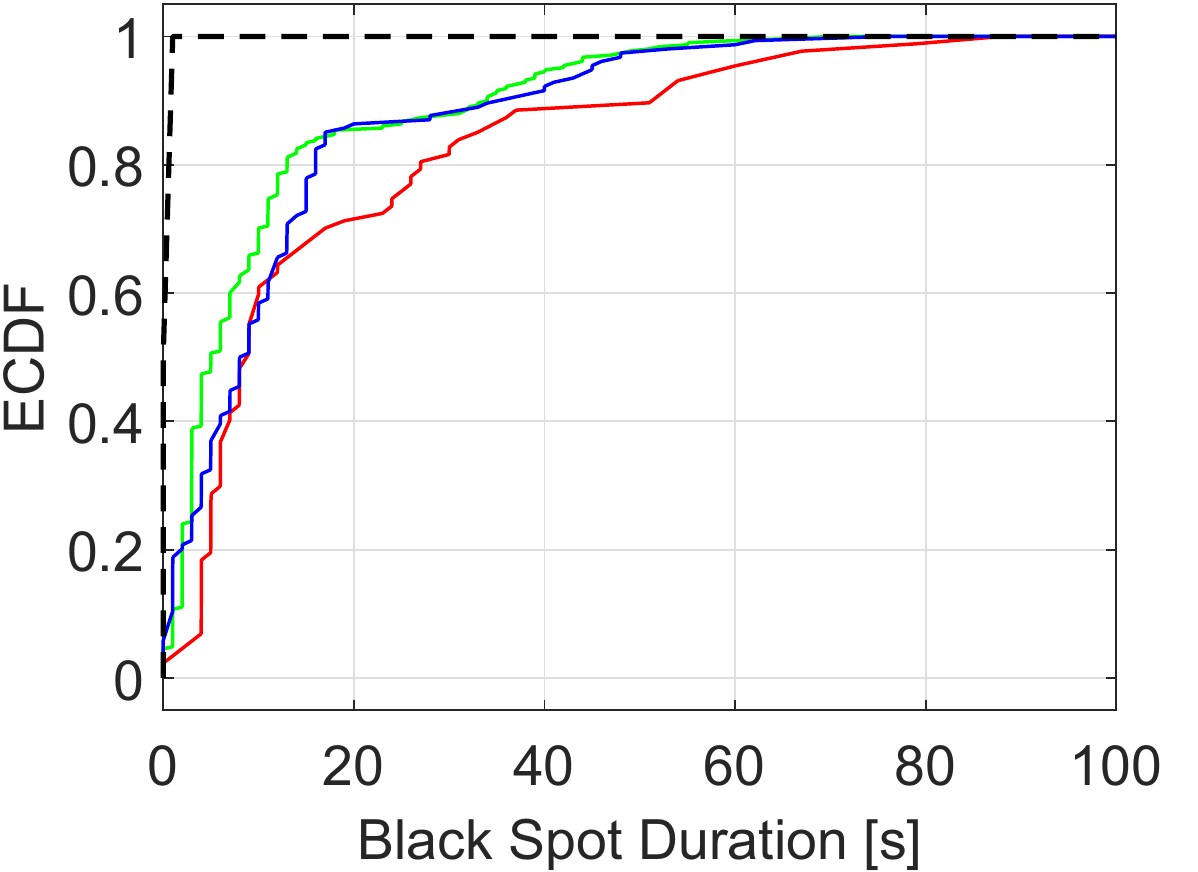}}\hfill		
	\caption{Compensation of black spot regions through multi-\ac{MNO} network selection: General solution approach and impact on black spot statistics.}
	\label{fig:black_spot_statistics}
	\vspace{-0.5cm}
\end{figure*}

%
% Phases
%
\begin{enumerate}
	%
	% Pre-trained
	%
	\item \textbf{Pre-trained model}: As the prediction model is initially optimized for being applied in the network of \mno{A}, the prediction accuracy for \mno{A} is significantly higher than for \mno{B}. Still, a certain level of predictability is achieved based on the \ac{MNO}-independent aspects within the feature set.
	%
	% Concept drift
	%
	\item \textbf{Concept drift}: After the first batches of \mno{B} measurements arrive, the prediction model experiences a concept drift: Since the weights of the \ac{ANN} are neither optimized for \mno{A} nor for \mno{B}, both models suffer from a performance decrease. Hereby, also the \ac{MNO}-independent features are affected from the changed model weights. This aspect is more dominant for \mno{B} for which only a small amount of measurements has been observed.
	%
	% Self adaptation
	%
	\item \textbf{Self adaptation}: After seven batch iterations, the \ac{ANN} weights start to become optimized for the network of \mno{B}, which results in a steady \ac{RMSE} improvement for the following iterations.
	%
	% Convergence
	%
	\item \textbf{Convergence}: After around 23 batch iterations, the prediction model reaches a converged state where the \ac{RMSE} stays at a nearly constant level. In comparison to the pre-trained phase, it can be seen that the \ac{RMSE} values of the two \acp{MNO} have been switched and that the model has successfully adopted itself for \mno{B}.
\end{enumerate}
%
% Conclusion: Online learning
%
The considered evaluation shows that online learning allows the data rate prediction model to autonomously adapt to changed network conditions which have a significant impact on the interplay of the features of the prediction model. Within the considered evaluation, even the on-device training time --- on average 0.4511~ms per 32-element batch --- can be considered negligible.
However, the considered \ac{ANN} model does not reach the accuracy level of the statically trained \ac{RF} predictor (see Fig. \ref{fig:rmse}). Therefore, future extensions should consider the application of more advanced methods for online learning.

%
% BLACK SPOT STATISTICS AND MULTI-MNO
%
\subsection{Black Spot Statistics and Multi-MNO Transmission Approach}

Although the previous discussion has shown that the black spot-aware data transfer approach is able to improve the data rate prediction accuracy as well as the resulting data rate of the \proposal method, it's usage introduces an additional buffering delay since transmissions are avoided within black spot regions. 
A possible solution approach for compensating these undesired effects might be the usage of a multi-\ac{MNO} approach which exploits complementary network infrastructure deployments. Fig.~\ref{fig:black_spot_statistics}~(a) shows a schematic visualization of black spot compensation through application of a multi-\ac{MNO} approach. If a vehicle encounters a black spot region within its primary network, it dynamically changes the network for performing the sensor data transmissions.

The \acp{ECDF} of the times and distances vehicles spend in black spot regions are shown in Fig.~\ref{fig:black_spot_statistics}~(b) and Fig.~\ref{fig:black_spot_statistics}~(c).
%
% MNOs
%
There are no significant variations between the considered \acp{MNO}. In around 50~\% of the cases, the black spot regions cover  less than 100~m which results in a minor addition to the buffering delay.
%
% Multi-MNO
%
The usage of a multi-\ac{MNO} approach leads to massive reductions of both undesired effects. In fact, it is also almost able to compensate the black spot-related effects completely.

\section{Recommendations for Future 6G Networks} \label{sec:recommendations}

Based on the achieved insights, we summarize the our recommendations for using client-based intelligence in future 6G networks as:
%
% Recommendations
%
\begin{itemize}
	%
	% Client-based approaches
	%
	\item \textbf{Non-cellular-centric networking} approaches such as end-edge-cloud orchestrated intelligence allow to exploit the computation and sensing capabilities of the network clients for participating in the overall network optimization. This potential should be recognized by the \acp{MNO} and actively supported.
	%
	% Cooperative prediction
	%
	\item Data rate prediction allows to make more precise statements about the channel quality than considering raw network quality indicators. Yet, purely client-based prediction methods only have limited insight into the current load of the network. As \textbf{cooperative data rate prediction} \cite{Sliwa/etal/2020a} is able to significantly reduce the end-to-end prediction error, this approach should be explicitly supported by the network infrastructure through actively sharing knowledge about the network load (e.g., obtained from the \ac{NWDAF} \cite{3GPP/2019a}) using dedicated control channel broadcasts.
	%
	% Data sets
	%
	\item Although machine learning has demonstrated its potential in various applications related to wireless network optimization, the sizes of most existing \textbf{data sets} are far away from being comparable to the massive data sets used in computer vision by industry giants. Therefore, effort should be taken to acquire data and build up massive open data sets, especially as additional data often leads to larger performance gains than model tuning \cite{Domingos/2012a}.
	%
	% Machine learning marketplace
	%
	A promising initial attempt for sharing data and models is the \emph{machine learning marketplace} proposed in draft recommendation Y.ML-IMT2020-MP of the \ac{ITU}.
\end{itemize}
\section{Conclusion} \label{sec:conclusion}

%
% Introduction
%
In this paper, we proposed \proposal as a novel method for resource-efficient opportunistic data transmission of vehicular sensor data.
%
% Solution appraoch
%
\proposal implements a hybrid machine learning approach which relies on supervised learning for data rate prediction, unsupervised learning for identifying geospatially-dependent uncertainties of the prediction model, and reinforcement learning for autonomously scheduling data transmissions with respect to the anticipated resource efficiency.
%
% Results
%
Within a real world performance evaluation campaign, it was shown that \proposal is able to achieve massive improvements in comparison to conventional periodic data transmission methods and significantly outperforms existing probabilistic approaches.
%
% Future work
%
In future work, we want to analyze more complex online learning approaches such as Mondrian Forest for the data rate prediction. In addition, our research work will focus on further improving the achievable prediction accuracy, e.g., through application of cooperative approaches.
\section*{Acknowledgment}

\footnotesize
Part of the work on this paper has been supported by Deutsche Forschungsgemeinschaft (DFG) within the Collaborative Research Center SFB 876 ``Providing Information by Resource-Constrained Analysis'', projects A4 and B4.

\vspace{-1.2cm}
\begin{IEEEbiography}[{\includegraphics[width=1in,height=1.25in,clip,keepaspectratio]{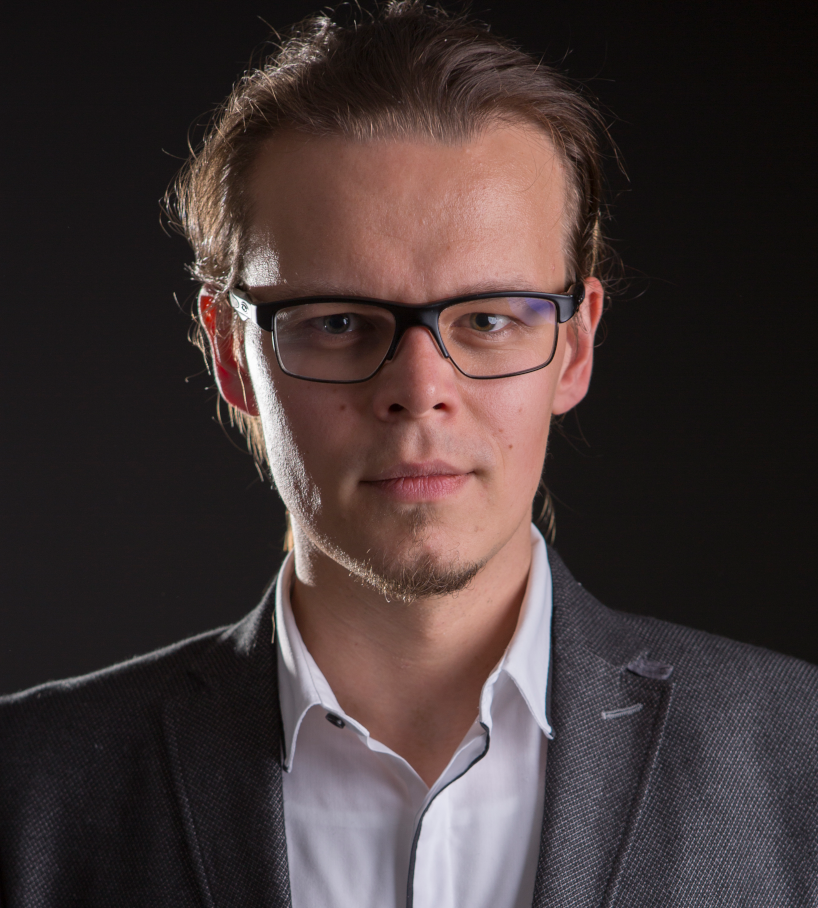}}]
	{Benjamin Sliwa}
	(S'16) received the M.Sc. degree from TU Dortmund University, Dortmund, Germany, in 2016. He is currently a Research Assistant with the Communication Networks Institute, Faculty of Electrical Engineering and Information Technology, TU Dortmund University. He is working on the Project "Analysis and Communication for Dynamic Traffic Prognosis" of the Collaborative Research Center SFB 876. His research interests include predictive and context-aware optimizations for decision processes in mobile and vehicular communication systems. Benjamin Sliwa has been recognized with a Best Paper Award at IEEE ICC 2020, a Best Student Paper Award at IEEE VTC-Spring 2018, the 2018 IEEE Transportation Electronics Student Fellowship "For Outstanding Student Research Contributions to Machine Learning in Vehicular Communications and Intelligent Transportation Systems", and a Best Contribution Award at the OMNeT++ Community Summit 2017.
\end{IEEEbiography}

\vspace{-1.2cm}
\begin{IEEEbiography}[{\includegraphics[width=1in,height=1.25in,clip,keepaspectratio]{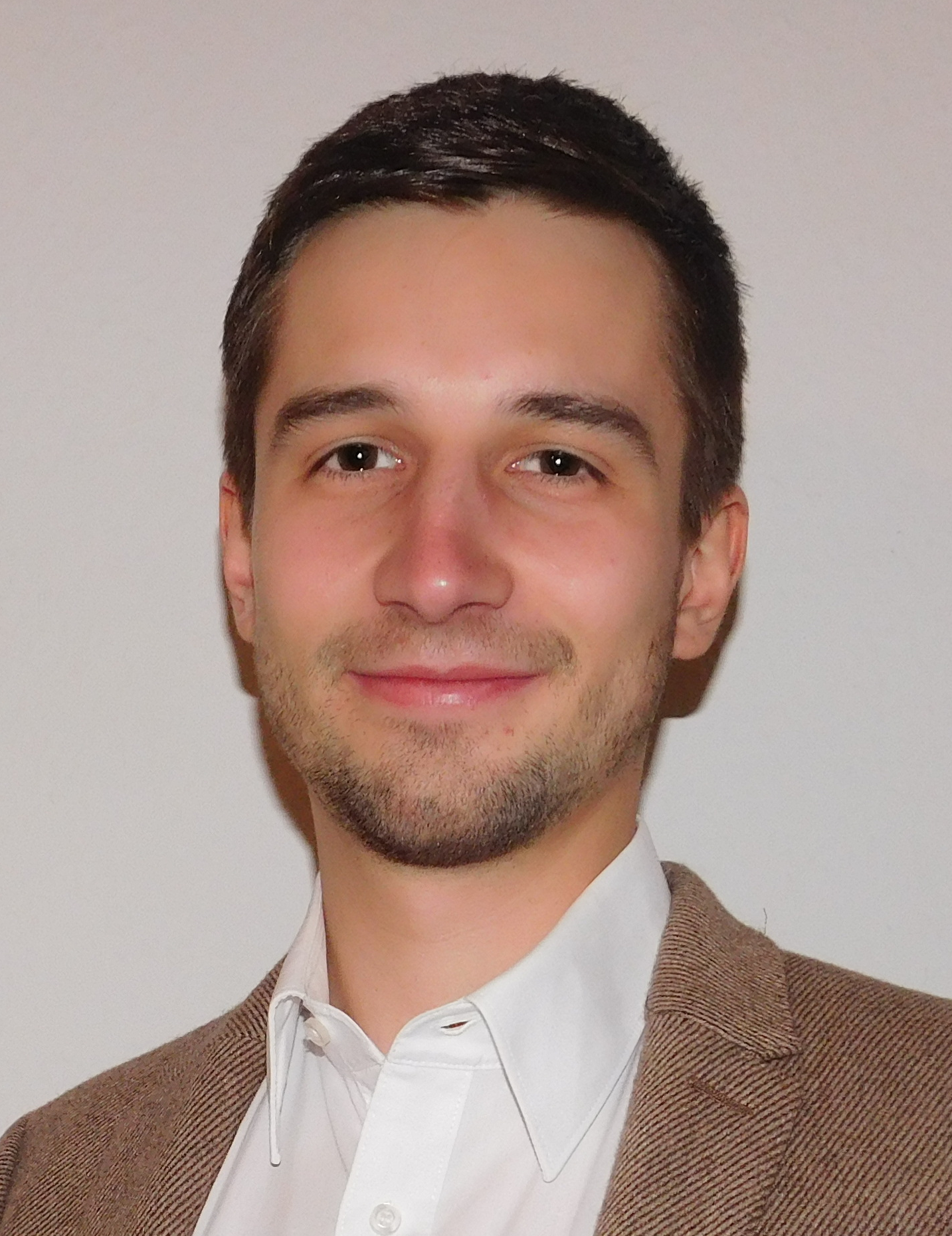}}]
{Rick Adam}
received the B.Sc. degree from TU Dortmund University, Dortmund, Germany, in 2017 and is currently working on his Master's Thesis at the Communication Networks Institute, Faculty of Electrical Engineering and Information Technology, TU Dortmund University. His research is focused on the application of machine learning algorithms for the optimization of communication networks, especially in the context of vehicular environments. One of the main goals is the development of a resource-efficient sensor data transmission system, which enables better coexistence between different applications in mobile networks.  
\end{IEEEbiography}

\vspace{-1.2cm}
\begin{IEEEbiography}
[{\includegraphics[width=1in,height=1.25in,clip,keepaspectratio]{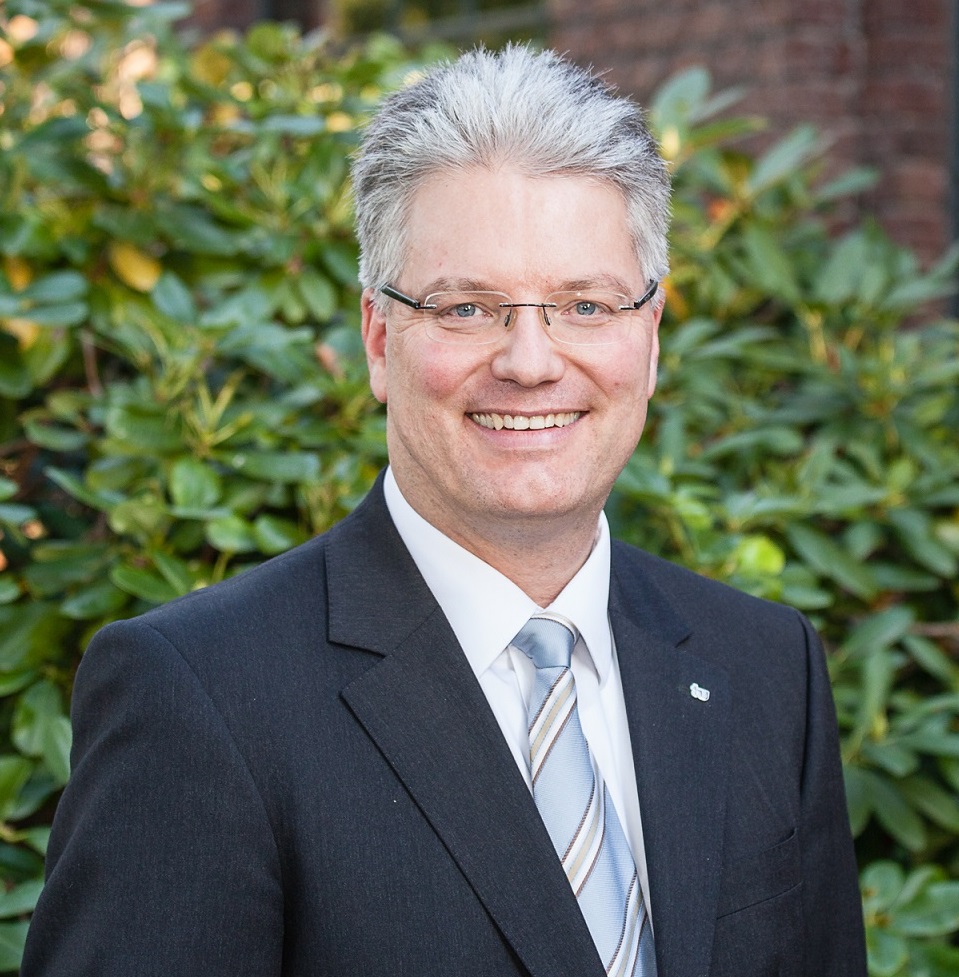}}]
{Christian Wietfeld}
(M’05–SM’12) received the Dipl.-Ing. and Dr.-Ing. degrees from RWTH Aachen University, Aachen, Germany.  He is currently a Full Professor of communication networks and the Head of the Communication Networks Institute, TU Dortmund University, Dortmund, Germany. For more than 20 years, he has been a coordinator of and a contributor to large-scale research projects on Internet-based mobile communication systems in academia (RWTH Aachen ‘92-’97, TU Dortmund since ‘05) and industry (Siemens AG ’97-’05). His current research interests include the design and performance evaluation of communication networks for cyber–physical systems in energy, transport, robotics, and emergency response.  He is the author of over 200 peer-reviewed papers and holds several patents. Dr. Wietfeld is a Co-Founder of the IEEE Global Communications Conference Workshop on Wireless Networking for Unmanned Autonomous Vehicles and member of the Technical Editor Board of the IEEE Wireless Communication Magazine. In addition to several best paper awards, he received an Outstanding Contribution award of ITU-T for his work on the standardization of next-generation mobile network architectures.

\end{IEEEbiography}

\bibliographystyle{IEEEtran}
\bibliography{Bibliography}

\begin{thebibliography}{10}
\providecommand{\url}[1]{#1}
\csname url@samestyle\endcsname
\providecommand{\newblock}{\relax}
\providecommand{\bibinfo}[2]{#2}
\providecommand{\BIBentrySTDinterwordspacing}{\spaceskip=0pt\relax}
\providecommand{\BIBentryALTinterwordstretchfactor}{4}
\providecommand{\BIBentryALTinterwordspacing}{\spaceskip=\fontdimen2\font plus
\BIBentryALTinterwordstretchfactor\fontdimen3\font minus
  \fontdimen4\font\relax}
\providecommand{\BIBforeignlanguage}[2]{{%
\expandafter\ifx\csname l@#1\endcsname\relax
\typeout{** WARNING: IEEEtran.bst: No hyphenation pattern has been}%
\typeout{** loaded for the language `#1'. Using the pattern for}%
\typeout{** the default language instead.}%
\else
\language=\csname l@#1\endcsname
\fi
#2}}
\providecommand{\BIBdecl}{\relax}
\BIBdecl

\bibitem{Xu/etal/2018a}
W.~{Xu}, H.~{Zhou}, N.~{Cheng}, F.~{Lyu}, W.~{Shi}, J.~{Chen}, and X.~{Shen},
  ``Internet of vehicles in big data era,'' \emph{IEEE/CAA Journal of
  Automatica Sinica}, vol.~5, no.~1, pp. 19--35, Jan 2018.

\bibitem{Ren/etal/2015a}
J.~{Ren}, Y.~{Zhang}, K.~{Zhang}, and X.~{Shen}, ``Exploiting mobile
  crowdsourcing for pervasive cloud services: {C}hallenges and solutions,''
  \emph{IEEE Communications Magazine}, vol.~53, no.~3, pp. 98--105, 2015.

\bibitem{Sliwa/etal/2019b}
B.~Sliwa, T.~Liebig, T.~Vranken, M.~Schreckenberg, and C.~Wietfeld,
  ``System-of-systems modeling, analysis and optimization of hybrid vehicular
  traffic,'' in \emph{2019 Annual IEEE International Systems Conference
  (SysCon)}, Orlando, Florida, USA, Apr 2019.

\bibitem{Akpakwu/etal/2018a}
G.~A. {Akpakwu}, B.~J. {Silva}, G.~P. {Hancke}, and A.~M. {Abu-Mahfouz}, ``A
  survey on {5G} networks for the internet of things: {C}ommunication
  technologies and challenges,'' \emph{IEEE Access}, vol.~6, pp. 3619--3647,
  2018.

\bibitem{Capponi/etal/2019a}
A.~{Capponi}, C.~{Fiandrino}, B.~{Kantarci}, L.~{Foschini}, D.~{Kliazovich},
  and P.~{Bouvry}, ``A survey on mobile crowdsensing systems: {C}hallenges,
  solutions, and opportunities,'' \emph{IEEE Communications Surveys Tutorials},
  vol.~21, no.~3, pp. 2419--2465, 2019.

\bibitem{AECC/2020a}
AECC, ``White paper: {O}perational behavior of a high definition map
  application,'' Automotive Edge Computing Consortium, Tech. Rep., May 2020.

\bibitem{3GPP/2019a}
3GPP, ``{3GPP TS 29.520 - 5G System; Network Data Analytics Services;Stage
  3},'' {3rd Generation Partnership Project (3GPP)}, Tech. Rep. 29.520, Mar
  2019, version 15.3.0.

\bibitem{Yang/etal/2019a}
P.~{Yang}, Y.~{Xiao}, M.~{Xiao}, and S.~{Li}, ``{6G} wireless communications:
  {Vi}sion and potential techniques,'' \emph{IEEE Network}, vol.~33, no.~4, pp.
  70--75, July 2019.

\bibitem{Ali/etal/2020a}
S.~Ali, W.~Saad, N.~Rajatheva, K.~Chang, D.~Steinbach, B.~Sliwa, C.~Wietfeld,
  K.~Mei, H.~Shiri, H.~Zepernick, T.~M.~C. Chu, I.~Ahmad, J.~Huusko,
  J.~Suutala, S.~Bhadauria, V.~Bhatia, R.~Mitra, S.~Amuru, R.~Abbas, B.~Shao,
  M.~Capobianco, G.~Yu, M.~Claes, T.~Karvonen, M.~Chen, M.~Girnyk, and
  H.~Malik, ``{6G} white paper on machine learning in wireless communication
  networks,'' Apr 2020.

\bibitem{Ren/etal/2019a}
J.~Ren, D.~Zhang, S.~He, Y.~Zhang, and T.~Li, ``A survey on end-edge-cloud
  orchestrated network computing paradigms: {T}ransparent computing, mobile
  edge computing, fog computing, and cloudlet,'' \emph{ACM Comput. Surv.},
  vol.~52, no.~6, Oct. 2019.

\bibitem{Sliwa/Wietfeld/2020a}
B.~Sliwa and C.~Wietfeld, ``A reinforcement learning approach for efficient
  opportunistic vehicle-to-cloud data transfer,'' in \emph{2020 IEEE Wireless
  Communications and Networking Conference (WCNC)}, Seoul, South Korea, Apr
  2020.

\bibitem{Sliwa/etal/2020b}
B.~Sliwa, R.~Adam, and C.~Wietfeld, ``Acting selfish for the good of all:
  {C}ontextual bandits for resource-efficient transmission of vehicular sensor
  data,'' in \emph{Proceedings of the ACM MobiHoc Workshop on Cooperative Data
  Dissemination in Future Vehicular Networks (D2VNet)}, Online, Oct 2020.

\bibitem{Wang/etal/2020a}
J.~{Wang}, C.~{Jiang}, H.~{Zhang}, Y.~{Ren}, K.~{Chen}, and L.~{Hanzo},
  ``Thirty years of machine learning: {T}he road to pareto-optimal wireless
  networks,'' \emph{IEEE Communications Surveys Tutorials}, pp. 1--1, 2020.

\bibitem{Jiang/etal/2017a}
C.~Jiang, H.~Zhang, Y.~Ren, Z.~Han, K.~C. Chen, and L.~Hanzo, ``Machine
  learning paradigms for next-generation wireless networks,'' \emph{IEEE
  Wireless Communications}, vol.~24, no.~2, pp. 98--105, April 2017.

\bibitem{Ye/etal/2018a}
H.~Ye, L.~Liang, G.~Y. Li, J.~Kim, L.~Lu, and M.~Wu, ``Machine learning for
  vehicular networks: {R}ecent advances and application examples,'' \emph{IEEE
  Vehicular Technology Magazine}, vol.~13, no.~2, pp. 94--101, June 2018.

\bibitem{Sun/etal/2019a}
Y.~{Sun}, M.~{Peng}, Y.~{Zhou}, Y.~{Huang}, and S.~{Mao}, ``Application of
  machine learning in wireless networks: {K}ey techniques and open issues,''
  \emph{IEEE Communications Surveys Tutorials}, vol.~21, no.~4, pp. 3072--3108,
  2019.

\bibitem{LeCun/etal/2015a}
Y.~LeCun, Y.~Bengio, and G.~Hinton, ``\BIBforeignlanguage{English (US)}{Deep
  learning},'' \emph{\BIBforeignlanguage{English (US)}{Nature}}, vol. 521, no.
  7553, pp. 436--444, 5 2015.

\bibitem{Breiman/2001a}
L.~Breiman, ``Random forests,'' \emph{Mach. Learn.}, vol.~45, no.~1, pp. 5--32,
  Oct. 2001.

\bibitem{Rasmussen/2004a}
C.~E. Rasmussen, \emph{Gaussian Processes in Machine Learning}.\hskip 1em plus
  0.5em minus 0.4em\relax Berlin, Heidelberg: Springer Berlin Heidelberg, 2004,
  pp. 63--71.

\bibitem{Arthur/Vassilvitskii/2007a}
D.~Arthur and S.~Vassilvitskii, ``{k-means++}: {T}he advantages of careful
  seeding,'' in \emph{In Proceedings of the 18th Annual ACM-SIAM Symposium on
  Discrete Algorithms}, 2007.

\bibitem{Gacanin/2019a}
H.~{Gacanin}, ``Autonomous wireless systems with artificial intelligence: {A}
  knowledge management perspective,'' \emph{IEEE Vehicular Technology
  Magazine}, pp. 1--1, 2019.

\bibitem{Sutton/Barto/2018a}
R.~S. Sutton and A.~G. Barto, \emph{Reinforcement learning: {A}n introduction},
  2nd~ed.\hskip 1em plus 0.5em minus 0.4em\relax The MIT Press, 2018.

\bibitem{Watkins/Dayan/1992a}
C.~J. C.~H. Watkins and P.~Dayan, ``Q-learning,'' \emph{Machine Learning},
  vol.~8, no.~3, pp. 279--292, May 1992.

\bibitem{Sevgican/etal/2020a}
S.~{Sevgican}, M.~{Turan}, K.~{Gökarslan}, H.~B. {Yilmaz}, and T.~{Tugcu},
  ``Intelligent network data analytics function in {5G} cellular networks using
  machine learning,'' \emph{Journal of Communications and Networks}, vol.~22,
  no.~3, pp. 269--280, 2020.

\bibitem{3GPP/2019b}
3GPP, ``{3GPP TR 23.791 - Study of Enablers for Network Automation for 5G},''
  {3rd Generation Partnership Project (3GPP)}, Tech. Rep., Jun 2019, v16.2.0.

\bibitem{Park/etal/2019a}
J.~{Park}, S.~{Samarakoon}, M.~{Bennis}, and M.~{Debbah}, ``Wireless network
  intelligence at the edge,'' \emph{Proceedings of the IEEE}, vol. 107, no.~11,
  pp. 2204--2239, Nov 2019.

\bibitem{Doerner/etal/2018a}
S.~{D{\"o}rner}, S.~{Cammerer}, J.~{Hoydis}, and S.~t.~{Brink}, ``Deep learning
  based communication over the air,'' \emph{IEEE Journal of Selected Topics in
  Signal Processing}, vol.~12, no.~1, pp. 132--143, Feb 2018.

\bibitem{Sliwa/Wietfeld/2019d}
B.~Sliwa and C.~Wietfeld, ``Data-driven network simulation for performance
  analysis of anticipatory vehicular communication systems,'' \emph{IEEE
  Access}, Nov 2019.

\bibitem{Sliwa/Wietfeld/2019c}
------, ``Towards data-driven simulation of end-to-end network performance
  indicators,'' in \emph{2019 IEEE 90th Vehicular Technology Conference
  (VTC-Fall)}, Honolulu, Hawaii, USA, Sep 2019.

\bibitem{Cavalcanti/etal/2018a}
E.~R. Cavalcanti, J.~A.~R. de~Souza, M.~A. Spohn, R.~C. d.~M. Gomes, and A.~F.
  B. F.~d. Costa, ``{VANETs}' research over the past decade: {O}verview,
  credibility, and trends,'' \emph{SIGCOMM Comput. Commun. Rev.}, vol.~48,
  no.~2, pp. 31--39, May 2018.

\bibitem{Bui/etal/2017a}
N.~Bui, M.~Cesana, S.~A. Hosseini, Q.~Liao, I.~Malanchini, and J.~Widmer, ``A
  survey of anticipatory mobile networking: Context-based classification,
  prediction methodologies, and optimization techniques,'' \emph{IEEE
  Communications Surveys \& Tutorials}, 2017.

\bibitem{Toufga/etal/2019a}
S.~Toufga, S.~Abdellatif, P.~Owezarski, T.~Villemur, and D.~Relizani,
  ``Effective prediction of {V2I} link lifetime and vehicle's next cell for
  software defined vehicular networks: {A} machine learning approach,'' in
  \emph{IEEE Vehicular Networking Conference (VNC)}, Los Angeles, USA, Dec
  2019.

\bibitem{Dalgkitsis/etal/2020a}
A.~{Dalgkitsis}, P.~{Mekikis}, A.~{Antonopoulos}, and C.~{Verikoukis}, ``Data
  driven service orchestration for vehicular networks,'' \emph{IEEE
  Transactions on Intelligent Transportation Systems}, pp. 1--10, 2020.

\bibitem{Coll-Perales/etal/2019a}
B.~{Coll-Perales}, J.~{Gozalvez}, and J.~L. {Maestre}, ``{5G} and beyond:
  {S}mart devices as part of the network fabric,'' \emph{IEEE Network},
  vol.~33, no.~4, pp. 170--177, July 2019.

\bibitem{Ha/etal/2012a}
S.~Ha, S.~Sen, C.~Joe-Wong, Y.~Im, and M.~Chiang, ``{TUBE}: {T}ime-dependent
  pricing for mobile data,'' in \emph{Proceedings of the ACM SIGCOMM 2012
  Conference on Applications, Technologies, Architectures, and Protocols for
  Computer Communication}, ser. SIGCOMM ’12.\hskip 1em plus 0.5em minus
  0.4em\relax New York, NY, USA: Association for Computing Machinery, 2012, p.
  247–258.

\bibitem{Shi/etal/2014a}
C.~Shi, K.~Joshi, R.~K. Panta, M.~H. Ammar, and E.~W. Zegura, ``{CoAST}:
  {C}ollaborative application-aware scheduling of last-mile cellular traffic,''
  in \emph{Proceedings of the 12th Annual International Conference on Mobile
  Systems, Applications, and Services}, ser. MobiSys ’14.\hskip 1em plus
  0.5em minus 0.4em\relax New York, NY, USA: Association for Computing
  Machinery, 2014, p. 245–258.

\bibitem{Chakraborty/etal/2013a}
A.~Chakraborty, V.~Navda, V.~N. Padmanabhan, and R.~Ramjee, ``Coordinating
  cellular background transfers using loadsense,'' in \emph{Proceedings of the
  19th Annual International Conference on Mobile Computing \& Networking}, ser.
  MobiCom '13.\hskip 1em plus 0.5em minus 0.4em\relax New York, NY, USA:
  Association for Computing Machinery, 2013, p. 63–74.

\bibitem{Lee/etal/2019a}
J.~Lee, J.~Lee, Y.~Im, S.~Dhawaskar~Sathyanarayana, P.~Rahimzadeh, X.~Zhang,
  M.~Hollingsworth, C.~Joe-Wong, D.~Grunwald, and S.~Ha, ``{CASTLE} over the
  air: {D}istributed scheduling for cellular data transmissions,'' in
  \emph{Proceedings of the 17th Annual International Conference on Mobile
  Systems, Applications, and Services}, ser. MobiSys '19.\hskip 1em plus 0.5em
  minus 0.4em\relax New York, NY, USA: ACM, 2019, pp. 417--429.

\bibitem{Sliwa/etal/2019d}
B.~Sliwa, R.~Falkenberg, T.~Liebig, N.~Piatkowski, and C.~Wietfeld, ``Boosting
  vehicle-to-cloud communication by machine learning-enabled context
  prediction,'' \emph{IEEE Transactions on Intelligent Transportation Systems},
  Jul 2019.

\bibitem{Nikolov/etal/2020a}
G.~Nikolov, M.~Kuhn, A.~McGibney, and B.-L. Wenning, ``Reduced complexity
  approach for uplink rate trajectory prediction in mobile networks,'' in
  \emph{2020 ISSC, 31st Irish Signals and Systems Conference}, Jun 2020.

\bibitem{Lee/etal/2020a}
J.~Lee, S.~Lee, J.~Lee, S.~D. Sathyanarayana, H.~Lim, J.~Lee, X.~Zhu,
  S.~Ramakrishnan, D.~Grunwald, K.~Lee, and S.~Ha, ``{PERCEIVE}: {D}eep
  learning-based cellular uplink prediction using real-time scheduling
  patterns,'' in \emph{Proceedings of the 18th International Conference on
  Mobile Systems, Applications, and Services}, ser. MobiSys ’20.\hskip 1em
  plus 0.5em minus 0.4em\relax New York, NY, USA: Association for Computing
  Machinery, 2020, p. 377–390.

\bibitem{3GPP/2018b}
3GPP, \emph{3GPP TS 36.213 - LTE; Evolved Universal Terrestrial Radio Access
  (E-UTRA); Physical layer procedures (Release 15)}, 3rd Generation Partnership
  Project Technical Specification, Rev. V15.2.0, Oct 2018.

\bibitem{Samba/etal/2017a}
A.~Samba, Y.~Busnel, A.~Blanc, P.~Dooze, and G.~Simon, ``Instantaneous
  throughput prediction in cellular networks: {W}hich information is needed?''
  in \emph{2017 IFIP/IEEE Symposium on Integrated Network and Service
  Management (IM)}, May 2017, pp. 624--627.

\bibitem{Herrera-Garcia/etal/2019a}
A.~{Herrera-Garcia}, S.~{Fortes}, E.~{Baena}, J.~{Mendoza}, C.~{Baena}, and
  R.~{Barco}, ``Modeling of key quality indicators for end-to-end network
  management: {P}reparing for {5G},'' \emph{IEEE Vehicular Technology
  Magazine}, vol.~14, no.~4, pp. 76--84, Dec 2019.

\bibitem{Riihijarvi/Mahonen/2018a}
J.~Riihijarvi and P.~Mahonen, ``Machine learning for performance prediction in
  mobile cellular networks,'' \emph{IEEE Computational Intelligence Magazine},
  vol.~13, no.~1, pp. 51--60, Feb 2018.

\bibitem{Zappone/etal/2020a}
A.~Zappone, M.~D. Renzo, and M.~Debbah, ``Wireless networks design in the era
  of deep learning: {M}odel-based, {AI}-based, or both?'' \emph{IEEE
  Communications Magazine}, 2020.

\bibitem{Jomrich/etal/2018a}
F.~Jomrich, A.~Herzberger, T.~Meuser, B.~Richerzhagen, R.~Steinmetz, and
  C.~Wille, ``Cellular bandwidth prediction for highly automated driving -
  {E}valuation of machine learning approaches based on real-world data,'' in
  \emph{Proceedings of the 4th International Conference on Vehicle Technology
  and Intelligent Transport Systems 2018}, no.~4.\hskip 1em plus 0.5em minus
  0.4em\relax SCITEPRESS, Mar 2018, pp. 121--131.

\bibitem{Sliwa/Wietfeld/2019b}
B.~Sliwa and C.~Wietfeld, ``Empirical analysis of client-based network quality
  prediction in vehicular multi-{MNO} networks,'' in \emph{2019 IEEE 90th
  Vehicular Technology Conference (VTC-Fall)}, Honolulu, Hawaii, USA, Sep 2019.

\bibitem{Domingos/2012a}
P.~Domingos, ``A few useful things to know about machine learning,''
  \emph{Commun. ACM}, vol.~55, no.~10, p. 78–87, Oct. 2012.

\bibitem{Akselrod/etal/2017a}
M.~Akselrod, N.~Becker, M.~Fidler, and R.~Luebben, ``{4G} {LTE} on the road -
  what impacts download speeds most?'' in \emph{2017 IEEE 86th Vehicular
  Technology Conference (VTC-Fall)}, Sep. 2017, pp. 1--6.

\bibitem{Sliwa/etal/2020a}
B.~Sliwa, R.~Falkenberg, and C.~Wietfeld, ``Towards cooperative data rate
  prediction for future mobile and vehicular {6G} networks,'' in \emph{2nd 6G
  Wireless Summit (6G SUMMIT)}, Levi, Finland, Mar 2020.

\bibitem{Ide/etal/2015a}
C.~{Ide}, B.~{Dusza}, and C.~{Wietfeld}, ``Client-based control of the
  interdependence between {LTE} {MTC} and human data traffic in vehicular
  environments,'' \emph{IEEE Transactions on Vehicular Technology}, vol.~64,
  no.~5, pp. 1856--1871, 2015.

\bibitem{Sliwa/etal/2018b}
B.~Sliwa, T.~Liebig, R.~Falkenberg, J.~Pillmann, and C.~Wietfeld, ``Efficient
  machine-type communication using multi-metric context-awareness for cars used
  as mobile sensors in upcoming {5G} networks,'' in \emph{2018 IEEE 87th
  Vehicular Technology Conference (VTC-Spring)}, Porto, Portugal, Jun 2018,
  {Best student paper award}.

\bibitem{Zhou/etal/2019a}
Z.~{Zhou}, X.~{Chen}, E.~{Li}, L.~{Zeng}, K.~{Luo}, and J.~{Zhang}, ``Edge
  intelligence: {P}aving the last mile of artificial intelligence with edge
  computing,'' \emph{Proceedings of the IEEE}, vol. 107, no.~8, pp. 1738--1762,
  2019.

\bibitem{Li/etal/2010a}
L.~Li, W.~Chu, J.~Langford, and R.~E. Schapire, ``A contextual-bandit approach
  to personalized news article recommendation,'' in \emph{Proceedings of the
  19th International Conference on World Wide Web}, ser. WWW ’10.\hskip 1em
  plus 0.5em minus 0.4em\relax New York, NY, USA: Association for Computing
  Machinery, 2010, p. 661–670.

\bibitem{Satoda/etal/2020a}
K.~Satoda, E.~Takahashi, T.~Onishi, T.~Suzuki, D.~Ohta, K.~Kobayashi, and
  T.~Murase, ``Passive method for estimating available throughput for
  autonomous off-peak data transfer,'' \emph{Wireless Communications and Mobile
  Computing}, vol. 2020, pp. 1--12, 02 2020.

\bibitem{Falkenberg/etal/2018a}
R.~Falkenberg, B.~Sliwa, N.~Piatkowski, and C.~Wietfeld, ``Machine learning
  based uplink transmission power prediction for {LTE} and upcoming {5G}
  networks using passive downlink indicators,'' in \emph{2018 IEEE 88th
  Vehicular Technology Conference (VTC-Fall)}, Chicago, USA, Aug 2018.

\bibitem{Sliwa/etal/2020c}
B.~Sliwa, N.~Piatkowski, and C.~Wietfeld, ``{LIMITS}: {L}ightweight machine
  learning for {IoT} systems with resource limitations,'' in \emph{2020 IEEE
  International Conference on Communications (ICC)}, Dublin, Ireland, Jun 2020,
  {Best} paper award.

\bibitem{Hall/etal/2009a}
M.~Hall, E.~Frank, G.~Holmes, B.~Pfahringer, P.~Reutemann, and I.~H. Witten,
  ``The {WEKA} data mining software: {A}n update,'' \emph{SIGKDD Explorations},
  vol.~11, no.~1, pp. 10--18, 2009.

\bibitem{Gama/etal/2014a}
J.~a. Gama, I.~\v{Z}liobaite, A.~Bifet, M.~Pechenizkiy, and A.~Bouchachia, ``A
  survey on concept drift adaptation,'' \emph{ACM Comput. Surv.}, vol.~46,
  no.~4, Mar. 2014.

\bibitem{Lakshminarayanan/etal/2014a}
B.~Lakshminarayanan, D.~M. Roy, and Y.~W. Teh, ``Mondrian forests: {E}fficient
  online random forests,'' in \emph{Proceedings of the 27th International
  Conference on Neural Information Processing Systems - Volume 2}, ser.
  NIPS’14.\hskip 1em plus 0.5em minus 0.4em\relax Cambridge, MA, USA: MIT
  Press, 2014, p. 3140–3148.

\end{thebibliography}

\ifaccess
	\EOD
\fi

\end{document}